\documentclass[10pt]{article}
\usepackage[utf8]{inputenc}
\usepackage{graphicx}
\usepackage{float}
\usepackage{amsmath,amssymb}
\usepackage{hyperref}
\usepackage[T1]{fontenc}
\usepackage[backend=bibtex,style=numeric,natbib=true,sorting=none]{biblatex} 
\usepackage{geometry}
\newcommand{\U}{\Upsilon_b}
\newcommand{\G}{\Gamma}

\title{Exact solution of Kerr black hole perturbations via CFT$_2$ and instanton counting. \\ Greybody factor, Quasinormal modes and Love numbers}

\author{Giulio Bonelli\footnote{bonelli@sissa.it}, Cristoforo Iossa\footnote{ciossa@sissa.it}, Daniel Panea Lichtig\footnote{daniel.panea@sissa.it}, Alessandro Tanzini\footnote{tanzini@sissa.it}\\ \, \\
\normalsize International School of Advanced Studies (SISSA) via Bonomea 265, 34136 Trieste, Italy \\ \normalsize INFN, Sezione di Trieste \\
\normalsize Institute for Geometry and Physics, IGAP, via Beirut 2, 34136 Trieste, Italy}
\date{}

\begin{document}

\maketitle

\begin{abstract} 
We give explicit expressions for the finite frequency greybody factor, quasinormal modes and Love numbers of Kerr black holes by computing the exact connection coefficients of the radial and angular parts of the Teukolsky equation. 
This is obtained by solving the connection problem of the \textit{confluent} Heun
equation in terms of the explicit expression of irregular Virasoro conformal blocks as sums over partitions via the AGT correspondence.
In the relevant approximation limits our results are in agreement with existing literature. The method we use can be extended to solve the
linearized Einstein equation in other interesting gravitational backgrounds.
\end{abstract}
\newpage

\tableofcontents

\section{Introduction and outlook}

The recent experimental verification of gravitational waves 
\cite{Abbott:2016blz}
renewed the interest in the theoretical studies of General Relativity and black hole physics. A particularly interesting aspect is the development of exact computational techniques to produce high precision tests of General Relativity equations. From this perspective, the study of exact solutions of differential equations rather than their approximate or numerical solutions is of paramount importance both to deepen our comprehension of physical phenomena and to reveal possible physical fine structure effects.

On the other hand, recent developments in the study of two-dimensional conformal field theories, their relation with supersymmetric gauge theories, equivariant localisation and duality in quantum field theory produced
new tools which are very effective to study long-standing classical problems in the theory of differential equations. Indeed, 
it has been known for a long time that the study of two-dimensional Conformal Field Theories \cite{Belavin:1984vu}
and of the representations of its infinite-dimensional symmetry algebra
provide exact solutions to partial differential equations in terms of 
conformal blocks and the appropriate fusion coefficients. 
The prototypical example is the null-state equation at level 2 for primary operators
of Virasoro algebra
which reduce, in the large central charge limit, to a Schr\"odinger-like equation with regular singularities, corresponding to a potential term with at most quadratic poles.
In this way one can engineer solutions of second-order linear differential equations of Fuchsian type by making use of the appropriate two dimensional CFT\footnote{
Our analysis is here limited - for the sake of presenting the general method - to second order linear differential equations, but all we say can be generalized to higher order equations by considering higher level degenerate field insertions, as already considered in \cite{Belavin:1984vu}.}.
While under the operator/state correspondence the vertex operators in the above construction correspond to primary (highest weight) states, one can insert more general irregular vertex operators corresponding to universal Whittaker states. 
The latter generate irregular singularities in the corresponding null-state equation and therefore allow engineering more general potentials with singularities of order higher than two.
Schematically, given a multi-vertex operator ${\cal O}_V(z_1,\ldots,z_N)$ satisfying the OPE
\begin{equation}\label{OPE}
T(z){\cal O}_V(z_1,\ldots,z_N)
\sim 
V(z;z_i){\cal O}_V(z_1,\ldots,z_N)
\quad \mathrm{as} \quad 
z\sim z_i
\end{equation}
one finds the corresponding level 2 null-state equation
\begin{equation}\label{scro}
[b^{-2}\partial_z^2+\sum_i V(z;z_i)]\Psi(z)=0 \quad \Psi(z)=\langle\Phi_{2,1}(z) {\cal O}_V(z_1,\ldots,z_N)\rangle
\end{equation}
satisfied by the correlation function of the multi-vertex and the level $2$
degenerate field $\Phi_{2,1}(z)$.
If the multi-vertex contains primary operators only, the OPE \eqref{OPE} and the potential
in \eqref{scro} contain at most quadratic poles, while the insertions of irregular vertices 
generate higher order singularities in $\sum_i V(z;z_i)$. Actually, $V(z;z_i)$ is a function in
$z$ and in differential operators with respect to the $z_i$. 
The dependence on the latter is specified by the semiclassical limit $b\to 0$ of Liouville CFT\footnote{This is not to be confused with the semiclassical approximation of the Schr\"odinger equation.}, corresponding to large Virasoro central charge $c\to\infty$. 
In this way, one finds a Schr\"odinger-like equation
\begin{equation}
    \epsilon_1^2 \frac{d^2 \Psi(z)}{dz^2} + V_{CFT}(z) \Psi(z) = 0 \, ,
    \label{eq:schroedinger}
\end{equation}
where $\epsilon_1$ is a parameter which stays finite in the large central charge limit and plays the r\^ole of the Planck constant.
The advantage of this approach is that the explicit solution of the connection problem on the $z$-plane for equation \eqref{eq:schroedinger} can be derived from  
the explicit computation of the full CFT$_2$ correlator \eqref{eq:full} 
and from its expansions in different intermediate channels.
A crucial ingredient to accomplish this program is a deep control on the analytic structure of regular and irregular Virasoro conformal blocks.
This has been recently obtained after the seminal AGT paper \cite{Alday_2010}, where 
conformal blocks of Virasoro algebra have been identified with 
concrete combinatorial formulae arising from equivariant instanton counting
in the context of ${\cal N}=2$ four-dimensional supersymmetric 
gauge theories \cite{Nekrasov:2002qd,Nekrasov:2003rj}.
The explicit solution of the instanton counting problem has been decoded in the CFT language in terms of overlap of universal Whittaker states in \cite{Gaiotto:2009ma,Marshakov_2010,Bonelli_2012,Gaiotto:2012sf}.

More precisely, the wave function $\Psi(z)$ corresponds to the insertion 
of a BPS surface observable
in the gauge theory path integral 
\cite{Alday_2010a}.
The specific case studied in this paper corresponds to a surface observable in the $SU(2)$ ${\cal N}=2$ gauge theory with $N_f=3$ fundamental hypermultiplets. The relevant gauge theory moduli space
in these cases is the one of \textit{ramified} instantons 
\cite{Kanno:2011fw}, with vortices localised on the surface defect, the $z$-variable providing the fugacity for the vortex counting. In the simplest cases the latter is indeed captured by hypergeometric functions \cite{Bonelli:2011fq}.

An important consequence of the AGT correspondence between CFT correlation functions and exact BPS partition functions in ${\cal N}=2$ four dimensional gauge theories has been 
the discovery of the so called "Kiev formula" in the theory of
Painlev\'e transcendents \cite{Gamayun_2013}, which established the latter to be a further class of special functions with an explicit combinatorial expression in terms of equivariant volumes of instanton moduli spaces \cite{Nekrasov:2002qd,Bruzzo_2003}. 
This correspondence between Painlev\'e and gauge theory has been extended to the full Painlev\'e confluence diagram in \cite{Bonelli_2017}, used in \cite{bonelli2021instantons}
to produce recurrence relations for instanton counting for general gauge groups
and studied in terms of blow-up equations in \cite{Grassi:2016nnt,Bershtein:2018zcz,Nekrasov:2020qcq}.
These results are related via the AGT correspondence to the $c=1$ limit of Liouville conformal field theory. On the other hand, it is well-known that a direct relation exists between the linear system associated to Painlev\'e VI equation and the Heun equation \cite{Slavyanov:2000:SF}. Further studies on this subject appeared recently in \cite{Litvinov:2013sxa,Piatek:2017fyn,Lisovyy:2021bkm,Motohashi:2021zyv}.
This perspective has been analyzed in the context of black hole physics in \cite{dacunha2015kerr,Carneiro_da_Cunha_2016,Carneiro_da_Cunha_2020,amado2021remarks}
where it was suggested that some physical properties of black holes, such as their greybody factor and quasinormal modes, can be studied in a particular regime in terms of Painlev\'e equations. Numerical checks appeared in \cite{Novaes_2019,daCunha:2021jkm}.
A decisive step forward about the quasinormal mode problem has been taken in 
\cite{aminov2020black}, where a
different approach making use of the Seiberg-Witten quantum curve of an appropriate supersymmetric gauge theory has been advocated
to justify their sprectrum and whose evidence was also supported by comparison with
numerical analysis of the gravitational equation
(see also \cite{Hatsuda:2020sbn,Hatsuda:2020iql} for further developments).
This view point has been further analysed
in \cite{chico2021}, where the context is widely
generalized to D-branes and other types of gravitational backgrounds in various dimensions. 
From the CFT$_2$ viewpoint, the gauge theoretical approach corresponds to the large Virasoro central charge limit recalled above.
It would be interesting to explore the relation between the $c=1$ and $c=\infty$ approaches
(see \cite{Bershtein:2021uts} for recent interesting developments).
Let us remark that in our view the CFT$_2$ framework is the suitable one to provide a physical explanation of the above described relations
among black hole physics and supersymmetric gauge theories. 

In this paper, for the sake of concreteness and with a specific application to the Kerr black hole problem in mind, we study equation \eqref{scro} for $N_f=3$ in the case of two regular and one irregular singularity of fourth order.
In Sect.\ref{two} we review the relativistic massless wave equation
in the Kerr black hole background, giving rise to the Teukolsky equation, whose solution can be obtained by separation of variables.
In Sect.\ref{three} we recall how both the radial and angular parts 
reduce, under an appropriate dictionary, to \eqref{scro} with an irregular singularity of order four at infinity and two regular singularities,
which is the \textit{confluent} Heun equation
\cite{ronveaux1995heun}.
We provide the explicit exact solution of the connection coefficients in Sect.\ref{four}.
The efficiency of the instanton expansion in the exact solution against the numerical integration is demonstrated 
by a detailed quantitative analysis in Subsect.\ref{plots}.

In Sect.\ref{five} we apply these results to Kerr black hole physics. 

We perform the study of 
the greybody factor of the Kerr black hole at finite frequency 
for which we give an exact formula.
This reduces to the well-known result of Maldacena and Strominger \cite{Maldacena_1997} in the zero frequency limit and in the semiclassical regime reproduces  
the results computed via standard WKB approximation in \cite{dumlu2020stokes}.

By using the explicit solution of the connection problem, we also provide a proof of the exact quantization of Kerr black hole quasinormal modes as proposed in \cite{aminov2020black}.
By solving the angular Teukolsky equation, 
we also prove the analogue dual quantization condition on the corresponding parameters of the spin-weighted spheroidal harmonics.

Finally, we discuss the use of the precise asymptotics of our solution to determine the tidal deformation profile
in the far away region of the Kerr black hole and compare it to recent results 
on the associated Love numbers 
in the static \cite{Le_Tiec_2021} and quasi-static \cite{chia2020tidal,charalambous2021vanishing} regimes. We observe
that our method naturally distinguishes the source and response terms
in the solution without needing analytic continuation in the angular momentum \cite{Kol:2011vg,LeTiec:2020spy} and provides an alternative regularization procedure for the 
computation of static Love numbers.

\vspace{1cm}

Let us discuss some selected open points and possible further developments. 
\begin{itemize}
\item
from the CFT$_2$ perspective, the equation \eqref{scro} arises in the semiclassical limit of Liouville field theory. An intriguing question to investigate is whether the quantum corrections in CFT$_2$ can have a physical interpretation in the black hole description. In principle, this could be related to quantum gravitational corrections or more generally
to some deviations from General Relativity, which will affect
the physical properties of the black hole's gravitational field.

\item 
Although in a very different circle of ideas, a link of holographic type
between CFT$_2$ and Kerr black hole physics emerged in the last years since \cite{Guica_2009}.
It would be very interesting to find whether the mathematical structure behind the solution of the Kerr black hole radiation problem we present in this letter could have a clear interpretation in the context of the Kerr/CFT correspondence.

\item A further possible application of the method presented in this paper
is the study of the physics of the last stages of coalescence of compact objects with the Zerilli function \cite{Zerilli:1971wd}, see \cite{Annulli:2021dkw} for 
recent developments. The corresponding potential displays a fifth order singularity which can be engineered with a higher irregular state, corresponding
to Argyres-Douglas SCFT in gauge theory \cite{Argyres:1995xn}.
Let us remark that the CFT$_2$ methods 
extend beyond the equivariant 
localisation results in gauge theory, making it possible to quantitatively study higher order singularities \cite{Bonelli_2012}.

\item Other black hole backgrounds can be analysed with methods similar 
 to the ones used in this paper. An important example is given by Kerr black hole solutions which asymptote to the (Anti-)de Sitter metric at infinity. These correspond to the Heun equation, which has four regular singularities on the Riemann sphere, and can be engineered from five-point correlators in Liouville CFT with four primary operator insertions and one level 2 degenerate field.
 This will provide explicit formulae for the corresponding connection problem and wave functions allowing for example to give an exact expression for the greybody factor
 studied in \cite{Gregory:2021ozs}.
 
 \item Our method can well be extended to other gravitational potentials studied to analyse possible deviations from GR
 with a modified quasinormal mode spectrum \cite{Ikeda:2021uvc} and 
 Love numbers \cite{Brustein:2021bnw}.
 
 \item The results we present are given as a perturbative series in the instanton counting parameter $\Lambda$, which, as we show from comparison with the numerical solution in Subsect.\ref{plots}, actually converges very efficiently. From the gauge theory reader's viewpoint let us notice that understanding how to extend our approach to the connection problem on the $\Lambda$ plane \cite{Lisovyy:2018mnj} would improve our
 understanding the strong coupling effects in gauge theory. Moreover, it could reveal to be useful for other applications in gravitational
problems.

\end{itemize}

\section{Perturbations of Kerr black holes}\label{two}
The Kerr metric describes the spacetime outside of a stationary, rotating black hole in asymptotically flat space. In Boyer-Lindquist coordinates it reads: 
\begin{equation}
\begin{aligned}
ds^2= & -\left(\frac{\Delta - \text{a}^2 \sin^2 \theta}{\Sigma} \right) dt^{2} + \frac{\Sigma}{\Delta} dr^{2} + \Sigma d\theta^{2} + \left(\frac{(r^2+ \text{a}^2)^2 - \Delta \text{a}^2 \sin^2 \theta}{\Sigma} \right) \sin^{2}\theta \ d\phi^{2} \\ & - \frac{2\text{a} \sin^{2} \theta (r^2 + \text{a}^2 -\Delta)}{\Sigma} dt \, d\phi \,,
\end{aligned}
\end{equation}
where
\begin{equation}
\Sigma = r^2 + \text{a}^2 \cos^2 \theta \, , \quad \Delta = r^2 - 2Mr + \text{a}^2 \,.
\end{equation}
The horizons are given by the roots of $\Delta$:
\begin{equation}
r_\pm = M \pm \sqrt{M^2 - \text{a}^2} \,.
\end{equation}
Two other relevant quantities are the Hawking temperature and the angular velocity at the horizon:
\begin{equation}
T_H = \frac{r_+-r_-}{8 \pi M r_+}\, , \quad \Omega = \frac{\text{a}}{2Mr_+} \,.
\end{equation}
Perturbations of the Kerr metric by fields of spin $s=0,-1,-2$ are described by the Teukolsky equation \cite{Teukolsky:1972my}, who found that an Ansatz of the form 
\begin{equation}
\Phi_s = e^{im\phi - i\omega t} S_{\lambda, s} (\theta, \text{a} \omega) R_s (r) \,.
\end{equation}
permits a separation of variables of the partial differential equation. One gets\footnote{Dropping the ${}_s$ subscript to ease the notation} the following equations for the radial and the angular part (see for example \cite{Berti_2009} eq.25):
\begin{equation}
\begin{aligned}
&\Delta \frac{d^2R}{dr^2}+(s+1)\frac{d\Delta}{dr}\frac{dR}{dr} + \left(\frac{K^2-2is(r-M)K}{\Delta} - \Lambda_{\lambda, s} + 4 i s \omega r\right)R = 0 \,, \\
&\partial_x (1 - x^2) \partial_x S_\lambda + \left[ (cx)^2 + \lambda + s - \frac{(m+sx)^2}{1-x^2} - 2 c s x \right] S_\lambda = 0 \,.
\end{aligned}
\end{equation}
Here $x=\cos\theta$, $c = \text{a} \omega$ and
\begin{equation}
K=(r^2+\text{a}^2)\omega - \text{a}m, \quad \Lambda_{\lambda, s} = \lambda + \text{a}^2 \omega^2 - 2 \text{a} m \omega \,.
\end{equation}
$\lambda$ has to be determined as the eigenvalue of the angular equation with suitable boundary conditions imposing regularity at $\theta=0,\pi$. In general no closed-form expression is known, but for small $\text{a}\omega$ it is given by $\lambda = \ell(\ell+1) - s(s+1) + \mathcal{O}(\text{a}\omega)$. We give a way to calculate it to arbitrary order in $\text{a}\omega$ in subsection \ref{angular_quantization}. \newline
For later purposes it is convenient to write both equations in the form of a Schr\"odinger equation. For the radial equation we define
\begin{equation}
\begin{aligned}
& z = \frac{r-r_-}{r_+ - r_-}\, , \quad \psi(z) = \Delta(r)^\frac{s+1}{2} R(r) \,.
\end{aligned}
\end{equation}
With this change of variables the inner and outer horizons are at $z=0$ and $z=1$, respectively, and $r\rightarrow \infty$ corresponds to $z \rightarrow \infty$. We obtain the differential equation
\begin{equation}
\frac{d^2 \psi(z)}{dz^2} + V_r(z)\psi(z) = 0 
\end{equation}
with potential
\begin{equation}\label{radial_potential}
V_r(z) = \frac{1}{z^2 (z-1)^2}\sum_{i=0}^4 \hat{A}^r_i z^i\,.
\end{equation}
The coefficients $\hat{A}^r_i$ depend on the parameters of the black hole and the frequency, spin and angular momentum of the perturbation. Their explicit expression is given in Appendix \ref{coefficients_potential}.\newline
For the angular part instead we define
\begin{equation}
z = \frac{1+x}{2} \,, \quad y(z) = \sqrt{1-x^2} \frac{S_\lambda}{2} \,.
\label{eq:angchangeofvar}
\end{equation}
After this change of variables, $\theta = 0$ corresponds to $z = 1$, and $\theta = \pi$ to $z = 0$. The equation now reads
\begin{equation}
\frac{d^2 y(z)}{dz^2} + V_{ang} (z) y(z) = 0 \,,
\end{equation}
with potential
\begin{equation}
V_{ang} (z) = \frac{1}{z^2 (z-1)^2} \sum_{i=0}^4 \hat{A}_i^\theta z^i\,. 
\end{equation}
Again, we give the explicit expressions of the coefficients $\hat{A}_i^\theta$ in Appendix \ref{coefficients_potential}. When written as Schr\"odinger equations, it is evident that the radial and angular equations share the same singularity structure. They both have two regular singular points at $z = 0, 1$ and an irregular singular point of Poincar\'e rank one at $z=\infty$. Such a differential equation is well-known in the mathematics literature as the confluent Heun equation \cite{ronveaux1995heun}. 
\section{The confluent Heun equation and conformal field theory}\label{three}
\subsection{The confluent Heun equation in standard form}
The confluent Heun equation (CHE) is a linear differential equation of second order with regular singularities at $z=0$ and 1, and an irregular singularity of rank 1 at $z=\infty$. In its standard form it is written as
\begin{equation}\label{CHE_standard}
    \frac{d^2w}{dz^2}+\left(\frac{\gamma}{z}+\frac{\delta}{z-1}+\epsilon\right)\frac{dw}{dz}+\frac{\alpha z-q}{z(z-1)}w=0\,.
\end{equation}
By defining $w(z)=P(z)^{-1/2}\psi(z)$ with $P(z)=e^{\epsilon z} z^\gamma (z-1)^\delta$, we can bring the standard form of the CHE into the form of a Schr\"odinger equation:
\begin{equation}\label{CHE_Schrodinger}
    \frac{d^2\psi(z)}{dz^2} + V_{Heun}(z) \psi(z) = 0
\end{equation}
where the potential is
\begin{equation}
    V_{Heun}(z) = \frac{1}{z^2 (z-1)^2}\sum_{i=0}^4 A^H_i z^i \,,
\end{equation}
with coefficients $A_i$ given in terms of the parameters of the standard form of the CHE by
\begin{equation}\label{HeunAs}
\begin{aligned}
& A^H_0 = \frac{\gamma(2-\gamma)}{4} \,, \\
& A^H_1 = q+\frac{\gamma}{2}(\gamma+\delta-\epsilon-2) \,, \\
& A^H_2 = -q-\alpha-\frac{\gamma^2}{4}+\frac{\delta}{2}-\frac{(\delta-\epsilon)^2}{4}+\frac{\gamma}{2}(1-\delta+2\epsilon) \,, \\
& A^H_3 = \alpha-\frac{\epsilon}{2}(\gamma+\delta-\epsilon) \,, \\
& A^H_4 = - \frac{\epsilon^2}{4} \,.
\end{aligned}
\end{equation}
\subsection{The confluent Heun equation as a BPZ equation}\label{HeunasBPZ}
In this section we work at the level of chiral conformal field theory/conformal blocks, which are fixed completely by the Virasoro algebra. Throughout this paper we work with conformal momenta related to the conformal weight by $\Delta = \frac{Q^2}{4}-\alpha^2$. The representation theory of the Virasoro algebra contains degenerate Verma modules of weight $\Delta_{r,s} = \frac{Q^2}{4}-\alpha_{r,s}^2$ with $\alpha_{r,s}=-\frac{br}{2}-\frac{s}{2b}$, where $Q=b+\frac{1}{b}$ and $b$ is related to the central charge as $c=1+6Q^2$. At level 2, the degenerate field $\Phi_{2,1}$ has weight $\Delta_{2,1}=-\frac{1}{2}-\frac{3}{4}b^2$ and satisfies the null-state equation
\begin{equation}
    (b^{-2}L_{-1}^2 + L_{-2})\cdot \Phi_{2,1}(z) = 0\,.
    \label{eq:degenerateannihilation}
\end{equation}
When this field is inserted in correlation functions, equation (\ref{eq:degenerateannihilation}) translates into a differential equation for the correlator called BPZ equation \cite{Belavin:1984vu}. Consider then the following conformal block with a degenerate field insertion, which by a slight abuse of notation we denote by
\begin{equation}\label{correlator_unnormalized}
    \Psi (z) := \langle \Delta, \Lambda_0, m_0 | \Phi_{2,1}(z) V_2(1) | \Delta_1 \rangle\,.
\end{equation}
$\Phi_{2,1}$ is the degenerate field mentioned above, $V_2(1)$ is a primary operator of weight $\Delta_2=\frac{Q^2}{4}-\alpha_2^2$ inserted at $z=1$ and $|\Delta_1\rangle$ is a primary state of weight $\Delta_1=\frac{Q^2}{4}-\alpha_1^2$ corresponding via the state-operator correspondence to the insertion of $V_1(0)$. The state $\langle \Delta, \Lambda_0, m_0 |$, called an irregular state of rank 1, is a more exotic kind of state, defined in \cite{Marshakov_2009} as:
\begin{equation}\label{irregular_state}
    \langle \Delta, \Lambda_0, m_0| = \sum_Y \sum_p \langle\Delta|L_{Y} m_0^{|Y|-2p} \Lambda_0^{|Y|} Q_{\Delta}^{-1}\big([2^p,1^{|Y|-2p}],Y\big)\,.
\end{equation}
The first sum runs over Young tableaux $Y$, $|Y|$ denotes the total number of boxes in the tableau and $Q$ is the Shapovalov form $Q_\Delta(Y,Y')=\langle\Delta|L_Y L_{-Y'}|\Delta'\rangle$. The notation $[2^p,1^{|Y|-2p}]$ refers to a Young tableau with $p$ columns of two boxes and $|Y|-2p$ columns of single boxes. $p$ then runs from $0$ to $|Y|/2$. All in all this implies the following relations, derived in \cite{Marshakov_2009}:
\begin{equation}
\begin{aligned}
    & \langle \Delta, \Lambda_0, m_0|L_0  = \bigg( \Delta + \Lambda_0 \frac{\partial}{\partial \Lambda_0} \bigg)\langle \Delta, \Lambda_0, m_0| \,, \\
    & \langle \Delta, \Lambda_0, m_0|L_{-1} = m_0 \Lambda_0 \langle \Delta, \Lambda_0, m_0| \,, \\
    & \langle \Delta, \Lambda_0, m_0|L_{-2}  = \Lambda_0^2 \langle \Delta, \Lambda_0, m_0| \,, \\
    & \langle \Delta, \Lambda_0, m_0|L_{-n}  = 0 \quad \mathrm{for} \,\,  n \geq 3 \,,
\end{aligned}
\end{equation}
so it is a kind of coherent state for the Virasoro algebra. The investigation of these kind of states in CFT was motivated by the AGT conjecture \cite{Alday_2010} according to which they are related to asymptotically free gauge theories \cite{Gaiotto:2009ma,Bonelli_2012,Gaiotto:2012sf}. Indeed, this state can be obtained by colliding two primary operators mimicking the decoupling of a mass in the gauge theory \cite{Gaiotto:2012sf, Marshakov_2009}. The result of the collision, understood as a scaling limit of an OPE, naturally has nonzero overlap with any Verma module. This gives a so-called Whittaker state \cite{Gaiotto:2012sf} \cite{Lisovyy:2018mnj} \cite{Nagoya_2015}, denoted by $\langle \Lambda_0, m_0 |$ that makes no reference to any Verma module and is completely characterized by the following action of the Virasoro generators:
\begin{equation}
\begin{aligned}\label{eq:Whittakerstate}
    & \langle \Lambda_0, m_0|L_0  = \Lambda_0 \frac{\partial}{\partial \Lambda_0} \langle \Lambda_0, m_0| \,, \\
    & \langle \Lambda_0, m_0|L_{-1} = m_0 \Lambda_0 \langle \Lambda_0, m_0| \,, \\
    & \langle \Lambda_0, m_0|L_{-2}  = \Lambda_0^2 \langle \Lambda_0, m_0| \,, \\
    & \langle \Lambda_0, m_0|L_{-n}  = 0 \quad \mathrm{for} \,\,  n \geq 3 \,,
\end{aligned}
\end{equation}
The state introduced here is the projection of a Whittaker state onto a specific Verma module. Indeed, inserting explicitly the projector gives back the series (\ref{irregular_state}):
\begin{equation}
    \begin{aligned}
    \langle \Delta, \Lambda_0, m_0 | := \Lambda_0^{-\Delta} \langle \Lambda_0, m_0| \sum_{Y,Y'} Q^{-1}_\Delta(Y,Y') L_{-Y'}|\Delta \rangle \langle \Delta|L_{Y} = \sum_Y \sum_p \langle\Delta|L_{Y} m_0^{|Y|-2p} \Lambda_0^{|Y|} Q_{\Delta}^{-1}\big([2^p,1^{|Y|-2p}],Y\big)\,,
    \end{aligned}
\end{equation}
where the overlap of the Whittaker state with a primary is defined as $\langle \Lambda_0,m_0|\Delta\rangle = \Lambda_0^\Delta$. This correlator satisfies the following BPZ equation (see Appendix \ref{CFT_calculations} for details):
\begin{equation}
\begin{aligned}
    0 =& \langle \Delta, \Lambda_0, m_0 | \big(b^{-2} \partial_z^2 + L_{-2}\cdot \big) \Phi_{2,1}(z) V_2(1) | V_1 \rangle = \\
    = & \bigg(b^{-2} \partial_z^2 - \frac{1}{z}\partial_z - \frac{1}{z} \frac{1}{z-1} \big(z\partial_z - \Lambda_0 \partial_{\Lambda_0} + \Delta_{2,1} + \Delta_2 + \Delta_1 - \Delta \big) + \frac{\Delta_2}{(z-1)^2} + \frac{\Delta_1}{z^2} + \frac{ m_0 \Lambda_0}{z} + \Lambda_0^2\bigg) \Psi(z)\,.
\end{aligned}
\end{equation}
We now take a double-scaling limit known as the Nekrasov-Shatashvili (NS) limit in the AGT dual gauge theory \cite{NEKRASOV_2010}, which corresponds to the semiclassical limit of large Virasoro central charge in the CFT. This amounts to introducing a new parameter $\hbar$, and sending $\epsilon_2 = \hbar b \to 0$, while keeping fixed 
\begin{equation}\label{new}
\begin{aligned}
   & \epsilon_1 = \hbar/b, \\
   &\hat{\Delta}=\hbar^2 \Delta, \, \hat{\Delta}_1=\hbar^2 \Delta_1,\, \hat{\Delta}_2=\hbar^2 \Delta_2, \\
   & \Lambda = 2i\hbar \Lambda_0,\,  m_3 = \frac{i}{2}\hbar m_0 \, .
\end{aligned}    
\end{equation}
  Furthermore, arguments from CFT \cite{Zamolodchikov_1996} and the AGT conjecture tell us that in this limit the correlator exponentiates and the $z$-dependence appears only at subleading order:
\begin{equation}
    \Psi (z) \propto \exp{ \frac{1}{\epsilon_1 \epsilon_2} \left( \mathcal{F}^{\mathrm{inst}}(\epsilon_1)+ \epsilon_2 \mathcal{W}(z;\epsilon_1) + \mathcal{O}(\epsilon_2^2) \right)}\,.
\end{equation}
Introducing the normalized wavefunction $\psi(z) = \lim_{\epsilon_2 \rightarrow 0} \Psi(z) /\langle \Delta, \Lambda_0, m_0 | V_2(1) | \Delta_1 \rangle$ and multiplying everything by $\hbar^2$, the BPZ equation in the NS limit becomes
\begin{equation}
\begin{aligned}
    & 0 = \bigg(\epsilon_1^2\partial_z^2 - \frac{1}{z} \frac{1}{z-1} \big(- \Lambda \partial_{\Lambda} \mathcal{F}^{\mathrm{inst}} + \hat{\Delta}_2 + \hat{\Delta}_1 -\hat{\Delta} \big) + \frac{\hat{\Delta}_2}{(z-1)^2} + \frac{\hat{\Delta}_1}{z^2} - \frac{m_3 \Lambda}{z} - \frac{\Lambda^2}{4}\bigg) \psi(z) \,.
\end{aligned}
\label{eq:schroedingerexp}
\end{equation}
All other terms vanish in the limit. It takes the form of a Schr\"odinger equation: 
\begin{equation}
    \epsilon_1^2 \frac{d^2 \psi(z)}{dz^2} + V_{CFT}(z) \psi(z) = 0
    \label{eq:schroedinger}
\end{equation}
with potential
\begin{equation}
    V_{CFT}(z) = \frac{1}{z^2 (z-1)^2}\sum_{i=0}^4 A_i z^i\,.
\end{equation}
Written in this form it is clear that the BPZ equation for this correlation function takes the form of the confluent Heun equation. Using conformal momenta instead of dimensions we write $\hat{\Delta}_i=\frac{1}{4}-a_i^2$, where we have used $\hat{\Delta}_i=\hbar^2\Delta_i$, $\hbar Q=\epsilon_1+\epsilon_2=\epsilon_1$ and defined $a_i := \hbar \alpha_i$. Defining furthermore $E:=a^2-\Lambda \partial_{\Lambda} \mathcal{F}^{\mathrm{inst}}$, the coefficients of the potential are
\begin{equation}
\begin{aligned}
& A_0 = \frac{\epsilon_1^2}{4}-a_1^2 \,, \\
& A_1 = -\frac{\epsilon_1^2}{4} + E + a_1^2 - a_2^2 -m_3 \Lambda \,, \\
& A_2 = \frac{\epsilon_1^2}{4} -E + 2 m_3 \Lambda - \frac{\Lambda^2}{4} \,, \\
& A_3 = -m_3 \Lambda +\frac{\Lambda^2}{2} \,, \\
& A_4 = - \frac{\Lambda^2}{4}\,.
\end{aligned}
\label{eq:CFTA's}
\end{equation}
Comparing with the coefficients $A^H_i$ of the CHE in (\ref{HeunAs}) and setting $\epsilon_1=1$ to match the coefficient of the second derivative, we can identify the parameters of the standard form with the parameters of the CFT as:
\begin{equation}\label{CHE_dictionary}
    \boxed{\begin{aligned}
    & \alpha = \theta'' \Lambda(1+\theta a_1+\theta' a_2+\theta'' m_3) \,, \\
    & \gamma = 1+2\theta a_1 \,, \\
    & \delta = 1+2\theta' a_2 \,, \\
    &\epsilon = \theta'' \Lambda \,, \\
    & q = E-\frac{1}{4} - (\theta a_1+\theta' a_2)^2 - (\theta a_1+\theta' a_2) + \theta'' \Lambda\left(\frac{1}{2}+\theta a_1-\theta'' m_3\right) \,,
    \end{aligned}}
\end{equation}
for any choice of signs $\theta,\theta',\theta'' = \pm 1$. These $8=2^3$ dictionaries reflect the symmetries of the equation, which is invariant independently under $a_1\to -a_1$, $a_2\to-a_2$ and $(m_3,\Lambda)\to-(m_3,\Lambda)$.
\subsection{The radial dictionary}
We see that the BPZ equation takes the same form as the radial and angular equations of the black hole perturbation equation if we set $\epsilon_1 = 1$. We will do this from now on. This implies $b=\hbar$. 
Comparing with the coefficients $\hat{A}^r_i$ we find the following eight dictionaries between the parameters of the radial equation in the black hole problem and the CFT:
\begin{equation}
\begin{aligned}
& E = \frac{1}{4} + \lambda + s(s+1)+ \text{a}^2 \omega^2 - 8M^2 \omega^2 - \left( 2M\omega^2 + i s \omega \right) (r_+-r_-) \,, \\
& a_1 = \theta \left(-i \frac{\omega - m \Omega}{4 \pi T_H} + 2 \mathrm{i} M \omega + \frac{s}{2} \right) \,, \\
& a_2 = \theta' \left( -i\frac{\omega - m \Omega}{4 \pi T_H} - \frac{s}{2} \right) \,, \\
& m_3 = \theta'' \left(-2 i M \omega + s \right) \,, \\
& \Lambda = -2 i \theta '' \omega (r_+ - r_-) \,,
\end{aligned}
\end{equation}
where $\theta,\theta',\theta'' = \pm 1$. We will make the following choice for the dictionary from now on:
\begin{equation}\label{radial_dictionary}
\boxed{\begin{aligned}
& E = \frac{1}{4} + \lambda + s(s+1) + \text{a}^2 \omega^2 - 8M^2 \omega^2 - \left(2M\omega^2 + i s \omega \right) (r_+-r_-) \,, \\
& a_1 = -i \frac{\omega - m \Omega}{4 \pi T_H} + 2 \mathrm{i} M \omega + \frac{s}{2} \,, \\
& a_2 = -i \frac{\omega - m \Omega}{4 \pi T_H} - \frac{s}{2} \,, \\
& m_3 = -2 i  M \omega + s \,, \\
& \Lambda = -2 i \omega (r_+ - r_-) \,,
\end{aligned}}
\end{equation}
which corresponds to $\theta = \theta' = \theta'' = +1$. Using AGT this dictionary gives the following masses in the gauge theory (see Appendix \ref{AppendixNekrasov} for details):
\begin{equation}
\begin{aligned}
&m_1 = a_1+a_2= -i \frac{\omega - m \Omega}{2 \pi T_H} + 2i M \omega \,, \\
&m_2 = a_2-a_1= -2 i M \omega - s \,, \\
& m_3 = -2 i  M \omega + s \,. 
\end{aligned}
\end{equation}
This is the same result as the one found in \cite{aminov2020black} except for a shift in $E$, which is due to a different definition of the $U(1)$-factor.
\subsection{The angular dictionary}
Comparing instead (\ref{eq:CFTA's}) with the $\hat{A}_i^\theta$ in (\ref{eq:angkerrA's}) we find the following eight dictionaries between the parameters of the angular equation in the black hole problem and the CFT:
\begin{equation}
\begin{aligned}
&E = \frac{1}{4} + c^2 + s(s+1) - 2 c s  + \lambda \,, \\
&a_1 = \theta \left(-\frac{m-s}{2} \right) \,, \\
&a_2 = \theta' \left( -\frac{m+s}{2} \right) \,, \\
&m_3 = -\theta'' s \,, \\
&\Lambda = \theta'' 4 c \,,
\end{aligned}
\end{equation}
where again $\theta,\theta',\theta'' = \pm 1$ and our choice from here on will be $\theta = \theta' = \theta'' = +1$, i.e.:
\begin{equation}
\boxed{\begin{aligned}
& E =  \frac{1}{4} + c^2 + s(s+1) - 2 c s  + \lambda \,, \\
& a_1 = -\frac{m-s}{2} \,, \\
& a_2 = -\frac{m+s}{2}\,, \\
& m_3 = - s \,, \\
& \Lambda = 4 c \,.
\end{aligned}}
\label{eq:angulardictionaryy}
\end{equation}
Using AGT this dictionary gives the following masses in the gauge theory (see Appendix \ref{AppendixNekrasov} for details):
\begin{equation}
\begin{aligned}
&m_1 =a_1+a_2= - m \,, \\
&m_2 =a_2-a_1= - s \,, \\
&m_3 = - s \,.
\end{aligned}
\end{equation}
Again we note the discrepancy with \cite{aminov2020black} due to the different $U(1)$-factor.

\section{The connection problem}\label{section:ConnectionProblem}
\label{four}
Exploiting crossing symmetry of Liouville correlation functions we can connect different asymptotic expansions of the solutions of BPZ equations around different field insertion points. The DOZZ formula can be obtained exploiting the known connection formulae for hypergeometric functions \cite{Teschner_1995, Nieri_2014}. Here we do the reverse, namely knowing the DOZZ formula we reconstruct connection formulae for irregular degenerate conformal blocks. Asymptotic expansions are computed via OPEs with regular and irregular insertions. To this end, we recall that the OPE of the degenerate field of our interest and a primary field reads \cite{Belavin:1984vu}:
\begin{equation}
\Phi_{2,1} (z, \bar{z}) V_{\alpha_i} (w, \bar{w}) = \sum_{\pm} \mathcal{C}_{\alpha_{2, 1}, \alpha_i}^{\alpha_{i \pm}} |z-w|^{2 k_{\pm}} \left( V_{\alpha_{i \pm}} (w, \bar{w}) + \mathcal{O}(|z - w|^2 ) \right) \,,
\label{eq:regularOPE}
\end{equation}
where $\alpha_{i \pm} := \alpha_i \pm \frac{-b}{2}$, and $k_\pm = \Delta_{\alpha_{i \pm}} - \Delta_{\alpha_i} - \Delta_{2,1}$ is fixed by the $L_0$ action. The OPE coefficient $\mathcal{C}_{\alpha_{2, 1}, \alpha_i}^{\alpha_{i \pm}}$ is computed in terms of DOZZ factors \cite{Dorn:1994xn} \cite{Zamolodchikov:1995aa} (see Appendix \ref{DOZZfactors}), namely
\begin{equation}
\mathcal{C}_{\alpha_{2, 1}, \alpha_i}^{\alpha_{i \pm}} = G^{-1}(\alpha_{i \pm}) C(\alpha_{i \pm}, \frac{-b-Q}{2}, \alpha_i) \,.
\end{equation}
The OPE with the irregular state is constrained by conformal symmetry, and the leading behavior is fixed by the action of $L_0, L_1, L_2$ instead of just $L_0$. The overall factors are again given in terms of DOZZ factors (see Appendix \ref{IrregularOPE}). One finds
\begin{equation}
\begin{aligned}
\langle \Delta_\alpha, \Lambda_0, \bar{\Lambda}_0, m_0 | \Phi_{2, 1} (z, \bar{z}) &=  \mathcal{C}^{\alpha_+}_{\alpha, \alpha_{2,1}} \displaystyle\left\lvert \sum_{\pm, k} \mathcal{A}_{\alpha_+, m_{0 \pm}} (\pm \Lambda)^{-\frac{1}{2} \pm m_3 + b \alpha_+} z^{\frac{1}{2} (bQ - 1 \pm 2 m_3)} e^{\pm \Lambda z/2} z^{-k} \langle \Delta_{\alpha_+}, \Lambda_0, m_{0 \pm}; k | \right\rvert^2 + \\ &+ \mathcal{C}^{\alpha_-}_{\alpha, \alpha_{2,1}} \displaystyle\left\lvert \sum_{\pm, k} \mathcal{A}_{\alpha_-, m_{0 \pm}} (\pm \Lambda)^{-\frac{1}{2} \pm m_3 - b \alpha_-} z^{\frac{1}{2} (bQ - 1 \pm 2 m_3)} e^{\pm \Lambda z/2} z^{-k} \langle \Delta_{\alpha_-}, \Lambda_0, m_{0 \pm}; k | \right\rvert^2\,.
\end{aligned}
\label{eq:ierrgope}
\end{equation}
Here the irregular state depending on $\Lambda_0, \bar{\Lambda}_0$ denotes the full (chiral$\otimes$antichiral) state, and the modulus squared of the chiral states (depending only on $\Lambda_0$) also has to be understood as a tensor product. The coefficients $\mathcal{A}$ are given by
\begin{equation}
\begin{aligned}
&\mathcal{A}_{\alpha_+, m_{0 +}} = \frac{\Gamma (1 - 2 b \alpha_+)}{\Gamma (\frac{1}{2} + m_3 - b \alpha_+)} \,, \, \,
\mathcal{A}_{\alpha_+, m_{0 -}} =  \frac{\Gamma (1 - 2 b \alpha_+)}{\Gamma (\frac{1}{2} - m_3 - b \alpha_+)}  \,, \\
&\mathcal{A}_{\alpha_-, m_{0 +}} = \frac{\Gamma (1 + 2 b \alpha_-)}{\Gamma (\frac{1}{2} + m_3 + b \alpha_-)} \,, \, \,
\mathcal{A}_{\alpha_-, m_{0 -}} =  \frac{\Gamma (1 + 2 b \alpha_-)}{\Gamma (\frac{1}{2} - m_3 + b \alpha_-)}  \,.
\end{aligned}
\label{eq:irrOPEcoeffreal}
\end{equation}
Since the results presented in this section are formulated purely in a CFT context, they will be written for finite $b$ unless otherwise specified.
\subsection{Connection formulae for the irregular 4 point function}
Let us consider the irregular correlator
\begin{equation}
\Psi (z, \bar{z}) = \langle \Delta_\alpha, \Lambda_0, \bar{\Lambda}_0, m_0 | \Phi_{2, 1} (z, \bar{z}) V_{\alpha_2} (1, \bar{1}) | \Delta_{\alpha_1} \rangle \,.
\label{eq:full}
\end{equation}
The physical, crossing symmetric correlator has to be built using the Whittaker state (\ref{eq:Whittakerstate}) introduced before which makes no reference to $\Delta_\alpha$. Here instead we use the state projected onto the Verma module $\Delta_\alpha$ which provides us with the explicit expression (\ref{irregular_state}). In particular, the $\Lambda_0 \to 0$ limit is simple: it is just a primary state with the usual normalization. In any case, we still expect (\ref{eq:full}) to be crossing symmetric and we will exploit this in what follows. In a forthcoming paper we will show that the result presented here is consistent with crossing symmetry of the physical correlator. The asymptotics of $\Psi$ for $z \sim 1, \infty$, respectively $t, u-$channels, are given by the OPEs. Due to crossing symmetry, the two expansions have to agree, therefore
\begin{equation}\label{eq:asymptotics}
    \Psi(z, \bar{z}) = K_{\alpha_{2+}, \alpha_{2+}}^{(t)} | f_{\alpha_{2+}}^{(t)} (z) |^2 + K_{\alpha_{2-}, \alpha_{2-}}^{(t)} | f_{\alpha_{2-}}^{(t)} (z) |^2 = K_{\alpha_+, \alpha_+}^{(u)} | f_{\alpha_+}^{(u)} (z) |^2 + K_{\alpha_-, \alpha_-}^{(u)} | f_{\alpha_-}^{(u)} (z) |^2 \,.
\end{equation}
where
\begin{equation}
\begin{aligned}\label{Ks}
&K_{\alpha_{2+}, \alpha_{2+}}^{(t)} = \mathcal{C}_{\alpha_{2,1} \alpha_2}^{\alpha_{2+}} C(\alpha, \alpha_{2+}, \alpha_1) \,, \, K_{\alpha_{2-}, \alpha_{2-}}^{(t)} = \mathcal{C}_{\alpha_{2,1} \alpha_2}^{\alpha_{2-}} C(\alpha, \alpha_{2-}, \alpha_1) \,, \\
&K_{\alpha_{+}, \alpha_{+}}^{(u)} = \mathcal{C}_{\alpha_{2,1} \alpha}^{\alpha_+} C(\alpha_+, \alpha_2, \alpha_1) \,, \, K_{\alpha_{-}, \alpha_{-}}^{(u)} = \mathcal{C}_{\alpha_{2,1} \alpha}^{\alpha_-} C(\alpha_-, \alpha_2, \alpha_1) \,,
\end{aligned}
\end{equation}
are the DOZZ factors for the two fusion channels in the $t$ and $u$-channel OPEs and
\begin{equation}
\begin{aligned}
&f_{\alpha_{2+}}^{(t)} (z) = \langle \Delta_\alpha, \Lambda_0, m_0 | V_{\alpha_{2+}} (1) | \Delta_{\alpha_1} \rangle (z-1)^{\frac{b Q + 2 b \alpha_2}{2}} \left(1+\mathcal{O}(z-1)\right)\,, \\
&f_{\alpha_{2-}}^{(t)} (z) = \langle \Delta_\alpha, \Lambda_0, m_0 | V_{\alpha_{2-}} (1) | \Delta_{\alpha_1} \rangle (z-1)^{\frac{b Q - 2 b \alpha_2}{2}}\left(1+\mathcal{O}(z-1)\right)\left(1+\mathcal{O}(z-1)\right) \,, \\
&f_{\alpha_+}^{(u)} (z) = \sum_{\pm} \langle \Delta_{\alpha_+}, \Lambda_0, m_{0 \pm} | V_{\alpha_2} (1) | \Delta_{\alpha_1} \rangle \mathcal{A}_{\alpha_+, m_{0 \pm}} e^{\pm \frac{\Lambda z}{2}} (\pm \Lambda)^{-\frac{1}{2} \pm m_3 + b \alpha_+} z^{\frac{1}{2} \left( bQ - 1 \pm 2 m_3 \right)}\left(1+\mathcal{O}(z^{-1})\right) \,, \\
&f_{\alpha_-}^{(u)} (z) =\sum_{\pm} \langle \Delta_{\alpha_-}, \Lambda_0, m_{0 \pm} | V_{\alpha_2} (1) | \Delta_{\alpha_1} \rangle \mathcal{A}_{\alpha_-, m_{0 \pm}} e^{\pm \frac{\Lambda z}{2}} (\pm \Lambda)^{-\frac{1}{2} \pm m_3 - b \alpha_-} z^{\frac{1}{2} \left( bQ - 1 \pm 2 m_3 \right)}\left(1+\mathcal{O}(z^{-1})\right) \,,
\end{aligned}
\label{HeunLambdaasymptotics}
\end{equation}
give the expansions of the conformal blocks in the two fusion channels of the $t$ and $u$-channels. Here and in the following the chiral correlators have to be understood as conformal blocks, we have extracted the DOZZ factors and they appear in (\ref{Ks}). Note that in line with the definition (\ref{irregular_state}), the irregular state contributes to the DOZZ factor the same as a regular state. Here, as noted in section \ref{HeunasBPZ}, $f_{\pm}^{(t,u)}$ in the NS limit are (up to a rescaling by one of the correlators, to keep them finite) the two linearly independent confluent Heun functions expanded around $1$ and $\infty$, respectively. We remark that due to the presence of the irregular singularity the $\alpha_\pm$ channels at infinity contribute with two different irregular states each, corresponding to $m_{0 \pm}$. This is consistent with the fact that the irregular state comes from the collision of two primary operators \cite{Gaiotto:2012sf}. The two expansions are related via a connection matrix $M$ by
\begin{equation}
f_i^{(t)} (z) = M_{ij} f_j^{(u)} (z) \,, \, \, i = \alpha_{2 \pm} \,, \, j = \alpha_\pm \,.
\label{eq:connectionansatz}
\end{equation}
This equation, combined with the requirement of crossing symmetry (\ref{eq:asymptotics}) gives the constraints
\begin{equation}
K_{ij}^{(t)} M_{ik} M_{jl} = K_{kl}^{(u)} \,.
\label{eq:connectionconstraints}
\end{equation}
Equations (\ref{eq:connectionconstraints}) give 3 quadratic equations for the 4 entries $M_{ij}$. Other constraints come from noticing that the $M_{ij}$ have to respect the symmetry under reflection of the momenta. The sign ambiguity inherent in the quadratic constraints (\ref{eq:connectionconstraints}) is resolved by imposing that for $\Lambda \to 0$ they reduce to the known hypergeometric connection matrix, since
\begin{equation}
\langle \Delta_\alpha, \Lambda_0, \bar{\Lambda}_0, m_0 | \Phi_{2, 1} (z, \bar{z}) V_{\alpha_2} (1, \bar{1}) | \Delta_{\alpha_1} \rangle \to \langle \Delta_\alpha | \Phi_{2, 1} (z, \bar{z}) V_{\alpha_2} (1, \bar{1}) | \Delta_{\alpha_1} \rangle \,, \, \text{as} \, \Lambda \to 0 \,,
\end{equation}
and conformal blocks of the regular degenerate 4 point functions are hypergeometric functions. This gives
\begin{equation}
\begin{aligned}
&M_{\alpha_{2+}, \alpha_+} = \frac{\Gamma (- 2 b \alpha) \Gamma (1 + 2 b \alpha_2)}{\Gamma (\frac{1}{2} + b (\alpha_1 + \alpha_2 - \alpha)) \Gamma (\frac{1}{2} + b (-\alpha_1 + \alpha_2 - \alpha))} \,, \\
&M_{\alpha_{2-}, \alpha_-} =  \frac{\Gamma (2 b \alpha) \Gamma (1 - 2 b \alpha_2)}{\Gamma (\frac{1}{2} + b (\alpha_1 - \alpha_2 + \alpha)) \Gamma (\frac{1}{2} + b (-\alpha_1 - \alpha_2 + \alpha))} \,, \\
&M_{\alpha_{2+}, \alpha_-} = \frac{\Gamma (2 b \alpha) \Gamma (1 + 2 b \alpha_2)}{\Gamma (\frac{1}{2} + b (\alpha_1 + \alpha_2 + \alpha)) \Gamma (\frac{1}{2} + b (-\alpha_1 + \alpha_2 + \alpha))} \,, \\
&M_{\alpha_{2-}, \alpha_+} =   \frac{\Gamma (-2 b \alpha) \Gamma (1 - 2 b \alpha_2)}{\Gamma (\frac{1}{2} + b (\alpha_1 - \alpha_2 - \alpha)) \Gamma (\frac{1}{2} + b (-\alpha_1 - \alpha_2 - \alpha))} \,.
\end{aligned}
\label{eq:irregconnectionmatrix}
\end{equation}
Note that $M_{ij}$ is given by the hypergeometric connection matrix even for finite $\Lambda$, since all $\Lambda$ corrections are encoded in the asymptotics of the functions (\ref{HeunLambdaasymptotics}). Proceeding in the same way we can find connection coefficients between $0, 1$. Using crossing symmetry we have
\begin{equation}
\Psi (z, \bar{z}) = K_{\alpha_{1+}, \alpha_{1+}}^{(s)} | f_{\alpha_{1+}}^{(s)} (z) |^2 + K_{\alpha_{1-}, \alpha_{1-}}^{(s)} | f_{\alpha_{1-}}^{(s)} (z) |^2 = K_{\alpha_{2+}, \alpha_{2+}}^{(t)} | f_{\alpha_{2+}}^{(t)} (z) |^2 + K_{\alpha_{2-}, \alpha_{2-}}^{(t)} | f_{\alpha_{2-}}^{(t)} (z) |^2 \,,
\label{eq:01asymptotics}
\end{equation}
where
\begin{equation}
\begin{aligned}
&f_{\alpha_{1+}}^{(s)} (z) \simeq \langle \Delta_\alpha, \Lambda_0, m_0 | V_{\alpha_2} (1) | \Delta_{\alpha_{1+}} \rangle z^{\frac{b Q + b \alpha_1}{2}} \,, \\
&f_{\alpha_{1-}}^{(s)} (z) \simeq \langle \Delta_\alpha, \Lambda_0, m_0 | V_{\alpha_2} (1) | \Delta_{\alpha_{1-}} \rangle z^{\frac{b Q - b \alpha_1}{2}} \,.
\end{aligned}
\end{equation}
Imposing again
\begin{equation}
f_i^{(s)} (z) = N_{ij} f_j^{(t)} (z) \,,
\label{eq:connectionansatz01}
\end{equation}
substituting (\ref{eq:connectionansatz01}) in (\ref{eq:01asymptotics}) and imposing that $f^{(s,t)}$ reduce to hypergeometric functions as $\Lambda \to 0$ we find (see Appendix \ref{DOZZfactors}) 
\begin{equation}
\begin{aligned}
&N_{\alpha_{1+}, \alpha_{2+}} = \frac{\Gamma (-2 b \alpha_2) \Gamma (1 + 2 b \alpha_1)}{\Gamma (\frac{1}{2} + b (\alpha_1 - \alpha_2 + \alpha)) \Gamma (\frac{1}{2} + b (\alpha_1 - \alpha_2 - \alpha))} \,, \\
&N_{\alpha_{1-}, \alpha_{2-}} = \frac{\Gamma (2 b \alpha_2) \Gamma (1 - 2 b \alpha_1)}{\Gamma (\frac{1}{2} + b (-\alpha_1 + \alpha_2 - \alpha)) \Gamma (\frac{1}{2} + b (-\alpha_1 + \alpha_2 + \alpha))} \,, \\
&N_{\alpha_{1+}, \alpha_{2-}} = \frac{\Gamma (2 b \alpha_2) \Gamma (1 + 2 b \alpha_1)}{\Gamma (\frac{1}{2} + b (\alpha_1 + \alpha_2 - \alpha)) \Gamma (\frac{1}{2} + b (\alpha_1 + \alpha_2 + \alpha))} \,, \\
&N_{\alpha_{1-}, \alpha_{2+}} = \frac{\Gamma (-2 b \alpha_2) \Gamma (1 - 2 b \alpha_1)}{\Gamma (\frac{1}{2} + b (-\alpha_1 - \alpha_2 + \alpha)) \Gamma (\frac{1}{2} + b (-\alpha_1 - \alpha_2 - \alpha))} \,.
\end{aligned}
\end{equation}
\subsection{AGT dual of irregular correlators and NS limit}
The irregular correlators appearing in the asymptotics of the functions (\ref{HeunLambdaasymptotics}) can be efficiently computed as Nekrasov partition functions thanks to the AGT correspondence \cite{Alday_2010}. In particular, the irregular conformal block is identified with \cite{Marshakov_2009}
\begin{equation}
\langle \Delta_\alpha, \Lambda_0, m_0 | V_{\alpha_2} (1) | \Delta_{\alpha_1} \rangle = \mathcal{Z}^{\mathrm{inst}} (\Lambda, a, m_1, m_2, m_3) \,,
\end{equation}
where $\mathcal{Z}^{\mathrm{inst}} (\Lambda, a, m_1, m_2, m_3)$ is the Nekrasov instanton partition function of $SU(2)$ $\mathcal{N}=2$ gauge theory in the $\Omega$-background (see Appendix \ref{AppendixNekrasov}). While the analysis in the last section was completely general, in order to apply the obtained results to the Teukolsky equation, one needs to take the NS limit $\epsilon_2\to0$, $\epsilon_1=1$ as discussed in section \ref{HeunasBPZ}. In this limit the correlators diverge, but rescaling the functions in (\ref{HeunLambdaasymptotics}) by one of the correlators, the resulting ratios are finite. In a slight abuse of notation, we write the connection coefficients in the NS limit as
\begin{equation}
\begin{aligned}
&M_{a_{2+}, a_+} = \frac{\Gamma (- 2 a) \Gamma (1 + 2 a_2)}{\Gamma (\frac{1}{2} + a_1 + a_2 - a) \Gamma (\frac{1}{2} - a_1 + a_2 - a)} \,, \\
&M_{a_{2-}, a_-} =  \frac{\Gamma (2 a) \Gamma (1 - 2 a_2)}{\Gamma (\frac{1}{2} + a_1 - a_2 + a) \Gamma (\frac{1}{2} - a_1 - a_2 + a)} \,, \\
&M_{a_{2+}, a_-} = \frac{\Gamma (2 a) \Gamma (1 + 2 a_2)}{\Gamma (\frac{1}{2} + a_1 + a_2 + a) \Gamma (\frac{1}{2} - a_1 + a_2 + a)} \,, \\
&M_{a_{2-}, a_+} =   \frac{\Gamma (-2 a) \Gamma (1 - 2 a_2)}{\Gamma (\frac{1}{2} + a_1 - a_2 - a) \Gamma (\frac{1}{2} - a_1 - a_2 - a)} \,,
\end{aligned}
\label{eq:irregconnectionmatrixnsoinf}
\end{equation}
and similarly 
\begin{equation}
\begin{aligned}
&N_{a_{1+}, a_{2+}} = \frac{\Gamma (-2 a_2) \Gamma (1 + 2 a_1)}{\Gamma (\frac{1}{2} + a_1 - a_2 + a) \Gamma (\frac{1}{2} + a_1 - a_2 - a)} \,, \\
&N_{a_{1-}, a_{2-}} = \frac{\Gamma (2 a_2) \Gamma (1 - 2 a_1)}{\Gamma (\frac{1}{2} - a_1 + a_2 - a) \Gamma (\frac{1}{2} - a_1 + a_2 + a)} \,, \\
&N_{a_{1+}, a_{2-}} = \frac{\Gamma (2 a_2) \Gamma (1 + 2 a_1)}{\Gamma (\frac{1}{2} + a_1 + a_2 - a) \Gamma (\frac{1}{2} + a_1 + a_2 + a)} \,, \\
&N_{a_{1-}, a_{2+}} = \frac{\Gamma (-2 a_2) \Gamma (1 - 2 a_1)}{\Gamma (\frac{1}{2} - a_1 - a_2 + a) \Gamma (\frac{1}{2} - a_1 - a_2 - a)} \,,
\end{aligned}
\label{eq:irregconnectionmatrixnsoone}
\end{equation}
where $a_i=\hbar \alpha_i=b \alpha_i$ for $\epsilon_1=\hbar/b=1$.
\subsection{Plots of the connection coefficients}\label{plots}
In the following we illustrate the power of the connection coefficients obtained above by comparing our analytical solution to the numerical one. Furthermore this illustrates how to evaluate the connection coefficients. For simplicity we focus on the connection problem between $z=0$ and $1$. The confluent Heun function $w(z)$ solving the CHE in standard form (\ref{CHE_standard}) can be expanded as a power series near $z=0$ as
\begin{equation}
    w(z)=1-\frac{q}{\gamma}z+\frac{\alpha\gamma + q(q-\gamma-\delta+\epsilon)}{2\gamma(\gamma+1)}z^2+\mathcal{O}(z^3)\,.
\end{equation}
We are interested in analytically continuing this series toward the other singular point at $z=1$. This problem is solved by our connection coefficients, we just need to identify the functions and parameters: in terms of the function $\psi(z)$ solving the CHE in Schr\"odinger form (\ref{CHE_Schrodinger}), we have around $z=0$:
\begin{equation}
    \psi(z) = e^{\epsilon z/2} z^{\gamma/2}(z-1)^{\delta/2}w(z) = z^{\frac{1}{2}+\theta a_1}\left(1+\mathcal{O}(z)\right) = \hat{f}^{(s)}_{\alpha_{1\theta}}(z) \,,
\end{equation}
where we have introduced the normalized s-channel function, related to the s-channel function defined before by $f^{(s)}_{\alpha_{1\theta}}(z)=\langle \Delta_\alpha, \Lambda_0, m_0 | V_{\alpha_{2}} (1) | \Delta_{\alpha_{1\theta}} \rangle \hat{f}^{(s)}_{\alpha_{1\theta}}(z)$. Similarly, we define the normalized t-channel function, related to the one defined before by $f^{(t)}_{\alpha_{2\theta'}}(z)=\langle \Delta_\alpha, \Lambda_0, m_0 | V_{\alpha_{2\theta'}} (1) | \Delta_{\alpha_1} \rangle \hat{f}^{(t)}_{\alpha_{2\theta'}}(z)$. It is a solution to the CHE given as a power series around the singular point $z=1$ which can be obtained by the Fr\"obenius method:
\begin{equation}
    \hat{f}^{(t)}_{\alpha_{2\theta'}}(z) = (1-z)^{\frac{1}{2}+\theta'a_2}\left(1 - \frac{1/4 - a_1^2 -a_2^2+E}{1+2\theta'a_2}(1-z)+\mathcal{O}((1-z)^2)\right) \,.
\end{equation}
The s- and t-channel solutions are related by $f_i^{(s)} = N_{ij} f_j^{(t)} $, with the coefficients $N_{ij}$ given before, which we now give more explicitly:
\begin{equation}
    \boxed{\begin{aligned}
    \hat{f}^{(s)}_{\alpha_{1\theta}}(z) & = \frac{\Gamma (-2 a_2) \Gamma (1 + 2 \theta a_1)}{\Gamma (\frac{1}{2} + \theta a_1 - a_2 + a)) \Gamma (\frac{1}{2} + \theta a_1 - a_2 - a))} \frac{\langle \Delta_\alpha, \Lambda_0, m_0 | V_{\alpha_{2+}} (1) | \Delta_{\alpha_1} \rangle}{\langle \Delta_\alpha, \Lambda_0, m_0 | V_{\alpha_{2}} (1) | \Delta_{\alpha_{1\theta}} \rangle} \hat{f}^{(t)}_{\alpha_{2+}}(z) + \\
    &+ \frac{\Gamma (2 a_2) \Gamma (1 + 2 \theta a_1)}{\Gamma (\frac{1}{2} + \theta a_1 + a_2 + a)) \Gamma (\frac{1}{2} + \theta a_1 + a_2 - a))} \frac{\langle \Delta_\alpha, \Lambda_0, m_0 | V_{\alpha_{2-}} (1) | \Delta_{\alpha_1} \rangle}{\langle \Delta_\alpha, \Lambda_0, m_0 | V_{\alpha_{2}} (1) | \Delta_{\alpha_{1\theta}} \rangle} \hat{f}^{(t)}_{\alpha_{2-}}(z)
    \end{aligned}}
\end{equation}
for $\theta=\pm$. A further complication arises from the fact that the parameter in the CHE is $E$, but in the connection formula the parameter $a$ appears which is related to $E$ in a nontrivial way and has to be obtained by inverting the Matone relation \cite{Matone_1995,Flume_2004} (see Appendix \ref{AppendixNekrasov}):
\begin{equation}
    E=a^2-\Lambda\partial_\Lambda \mathcal{F}^{\mathrm{inst}}\,.
\label{eq:matone}
\end{equation}
Everything has to be computed for general $\epsilon_1,\epsilon_2$ using Nekrasov formulae and then specialized to the NS limit by setting $\epsilon_1=1$ and taking the limit $\epsilon_2\to0$ in the end. To work consistently at one instanton one also needs to expand the Gamma functions since they contain $a$ which is given as an instanton expansion. We get
\begin{equation}
    \boxed{\begin{aligned}
    &\hat{f}^{(s)}_{\alpha_{1\theta}}(z) = \frac{\Gamma (-2 a_2) \Gamma (1 + 2 \theta a_1)}{\Gamma (\frac{1}{2} + \theta a_1 - a_2 + \sqrt{E})) \Gamma (\frac{1}{2} + \theta a_1 - a_2 - \sqrt{E}))} \hat{f}^{(t)}_{\alpha_{2+}}(z)\times \\ & \times \left[1-\left(\frac{\theta a_1 + a_2}{\frac{1}{2}-2E} + \frac{\frac{1}{4}-E+a_1^2-a_2^2}{\sqrt{E}\left(1-4E\right)}\big[\psi^{(0)}\big(\frac{1}{2}-\sqrt{E}+\theta a_1 -a_2\big)-\psi^{(0)}\big(\frac{1}{2}+\sqrt{E}+\theta a_1 -a_2\big)\big]\right) m_3 \Lambda 
    \right] + \\
    &+ \frac{\Gamma (2 a_2) \Gamma (1 + 2 \theta a_1)}{\Gamma (\frac{1}{2} + \theta a_1 + a_2 + \sqrt{E})) \Gamma (\frac{1}{2} + \theta a_1 + a_2 - \sqrt{E}))} \hat{f}^{(t)}_{\alpha_{2-}}(z)\times \\ & \times \left[1-\left(\frac{\theta a_1 - a_2}{\frac{1}{2}-2E}+ \frac{\frac{1}{4}-E+a_1^2-a_2^2}{\sqrt{E}\left(1-4E\right)}\big[\psi^{(0)}\big(\frac{1}{2}-\sqrt{E}+\theta a_1 +a_2\big)-\psi^{(0)}\big(\frac{1}{2}+\sqrt{E}+\theta a_1+a_2\big)\big] \right)m_3 \Lambda 
    \right]\\
    &+ \mathcal{O}(\Lambda^2).
    \end{aligned}}
\end{equation}
Here $\psi^{(0)}(z)=\frac{d}{dz}\log \Gamma(z)$ is the digamma function. The higher instanton corrections to the connection coefficients can be computed in an analogous way.
We have identified $\hat{f}^{(s)}_{\alpha_{1\theta}}(z) = e^{\epsilon z/2} z^{\gamma/2}(z-1)^{\delta/2}w(z)$ by using the power series expansion near $z=0$. We can then use the connection formula given above to obtain the power series expansion near $z=1$ in terms of $\hat{f}^{(t)}_{\alpha_{2\pm}}(z)$, and compare it to the numerical solution. In the following we illustrate the power of the connection formula by giving random values (in a suitable range) to the various parameters and plotting the confluent Heun function numerically versus the three-term power expansion at $z=1$, computed analytically by using the connection formula from $0$ to $1$. Here we use the dictionary between the parameters of the CHE in standard form and the CFT parameters given in (\ref{CHE_dictionary}), with $\theta = +1,\theta'=-1,\theta''=-1$.
\begin{figure}[H]
\centering
\includegraphics[width=\textwidth]{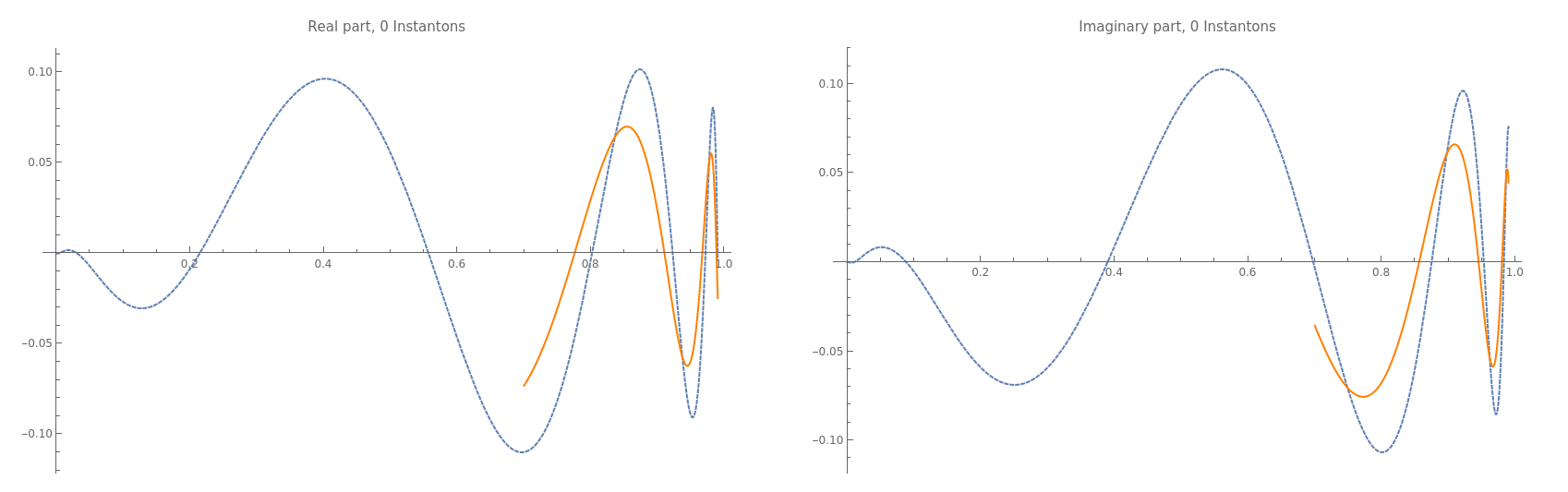}
\caption{{\small Real and imaginary parts of the rescaled confluent Heun function $e^{\epsilon z/2} z^{\gamma/2}(z-1)^{\delta/2}w(z)$ (blue, dashed), computed numerically, and of the three-term power expansion near $z=1$ (solid, orange), obtained analytically using the connection coefficients computed at zero instantons. The validity of the series expansion around $z=1$ (orange) is limited to a neighborhood of $z=1$, but going to higher orders in the expansion to extend the validity is straightforward. The values of the parameters are:
$a_1 = 0.970123 + 1.36981i,\, a_2 = -0.386424 - 2.99783i,\, E = 5.41627 + 6.40871i,\, m_3 = 1.68707 - 0.707722i,\,\Lambda = 1.96772 + 1.80414i$.}}
\end{figure}

\begin{figure}[H]
\centering
\includegraphics[width=\textwidth]{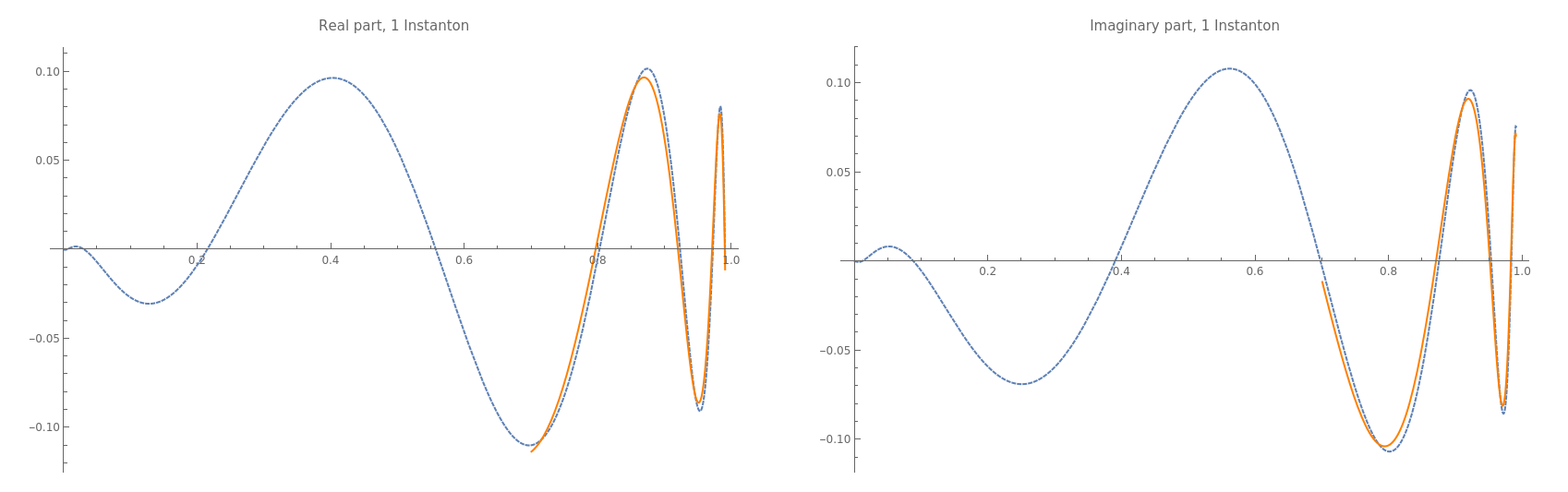}
\caption{{\small Real and imaginary parts of the rescaled confluent Heun function $e^{\epsilon z/2} z^{\gamma/2}(z-1)^{\delta/2}w(z)$ (blue, dashed), computed numerically, and of the three-term power expansion near $z=1$ (solid, orange), obtained analytically using the connection coefficients computed at one instanton. The validity of the series expansion around $z=1$ (orange) is limited to a neighborhood of $z=1$, but going to higher orders in the expansion to extend the validity is straightforward. The values of the parameters are:
$a_1 = 0.970123 + 1.36981i,\, a_2 = -0.386424 - 2.99783i,\, E = 5.41627 + 6.40871i,\, m_3 = 1.68707 - 0.707722i,\,\Lambda = 1.96772 + 1.80414i$.}}
\end{figure}

\begin{figure}[H]
\centering
\includegraphics[width=\textwidth]{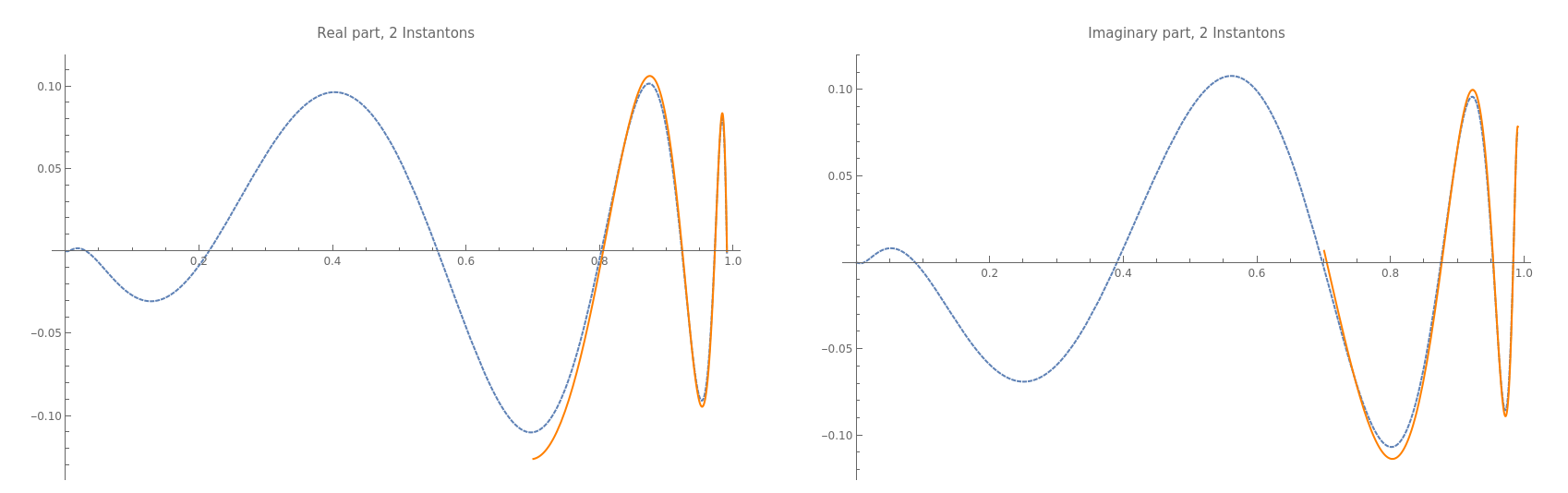}
\caption{{\small Real and imaginary parts of the rescaled confluent Heun function $e^{\epsilon z/2} z^{\gamma/2}(z-1)^{\delta/2}w(z)$ (blue, dashed), computed numerically, and of the three-term power expansion near $z=1$ (solid, orange), obtained analytically using the connection coefficients computed at two instantons. The validity of the series expansion around $z=1$ (orange) is limited to a neighborhood of $z=1$, but going to higher orders in the expansion to extend the validity is straightforward. The values of the parameters are:
$a_1 = 0.970123 + 1.36981i,\, a_2 = -0.386424 - 2.99783i,\, E = 5.41627 + 6.40871i,\, m_3 = 1.68707 - 0.707722i,\,\Lambda = 1.96772 + 1.80414i$.}}
\end{figure}

\begin{figure}[H]
\centering
\includegraphics[width=\textwidth]{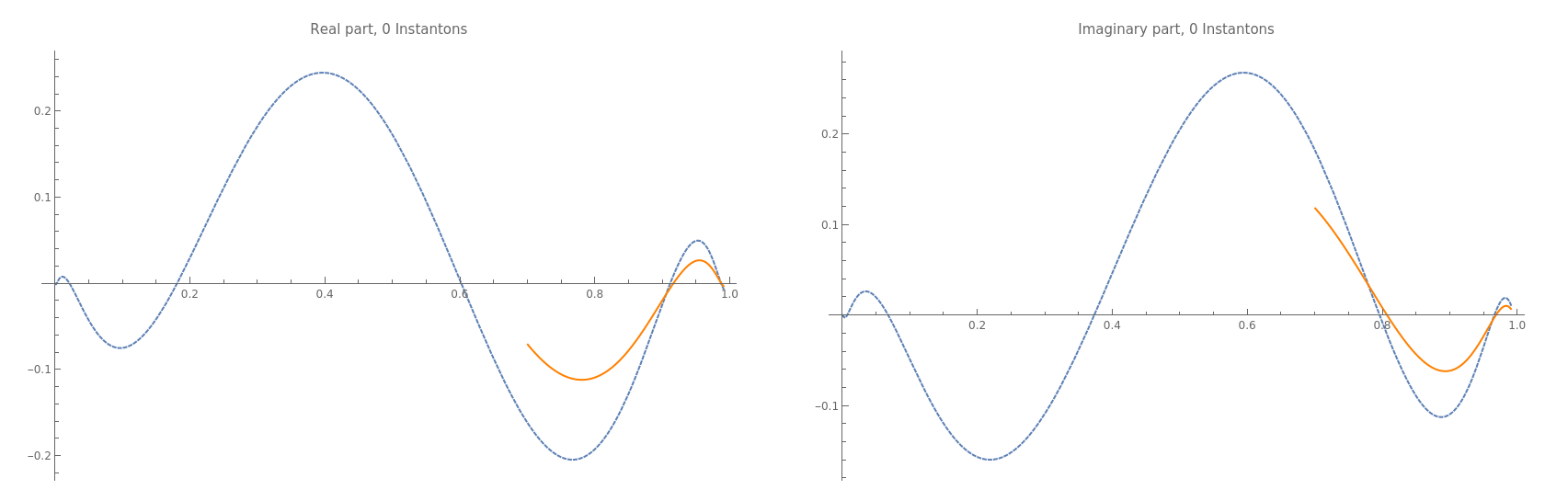}
\caption{{\small Real and imaginary parts of the rescaled confluent Heun function $e^{\epsilon z/2} z^{\gamma/2}(z-1)^{\delta/2}w(z)$ (blue, dashed), computed numerically, and of the three-term power expansion near $z=1$ (solid, orange), obtained analytically using the connection coefficients computed at zero instantons. The validity of the series expansion around $z=1$ (orange) is limited to a neighborhood of $z=1$, but going to higher orders in the expansion to extend the validity is straightforward. The values of the parameters are:
$a_1 = 0.5 + 1.24031i,\,a_2 = -0.5 + 1.55419i,\,E = 5.52396,\,m_3 = 0.92039 + 1.36765i,\, \Lambda=1.60238 + 1.25941i$.}}
\end{figure}

\begin{figure}[H]
\centering
\includegraphics[width=\textwidth]{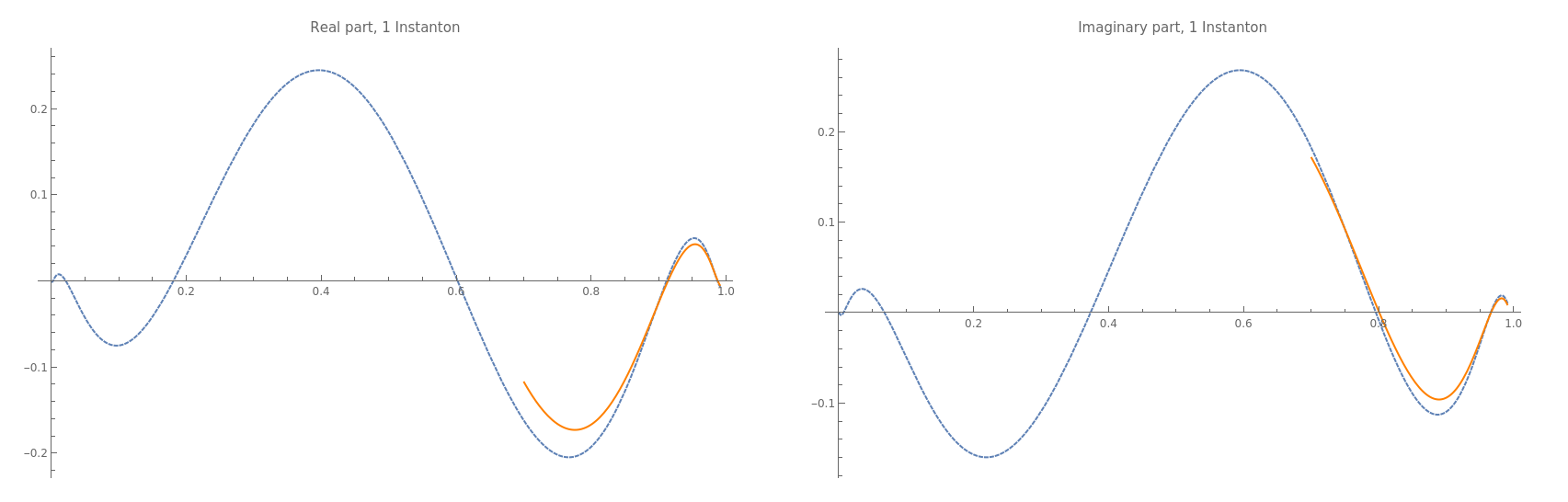}
\caption{{\small Real and imaginary parts of the rescaled confluent Heun function $e^{\epsilon z/2} z^{\gamma/2}(z-1)^{\delta/2}w(z)$ (blue, dashed), computed numerically, and of the three-term power expansion near $z=1$ (solid, orange), obtained analytically using the connection coefficients computed at one instanton. The validity of the series expansion around $z=1$ (orange) is limited to a neighborhood of $z=1$, but going to higher orders in the expansion to extend the validity is straightforward. The values of the parameters are:
$a_1 = 0.5 + 1.24031i,\,a_2 = -0.5 + 1.55419i,\,E = 5.52396,\,m_3 = 0.92039 + 1.36765i,\, \Lambda=1.60238 + 1.25941i$.}}
\end{figure}

\begin{figure}[H]
\centering
\includegraphics[width=\textwidth]{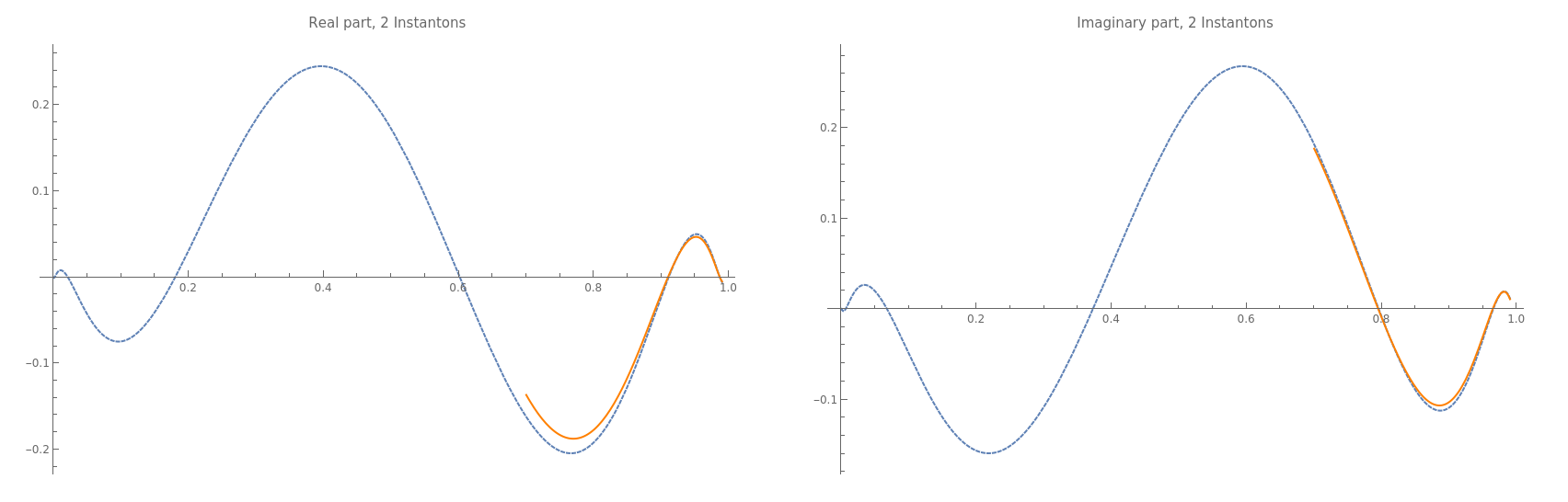}
\caption{{\small Real and imaginary parts of the rescaled confluent Heun function $e^{\epsilon z/2} z^{\gamma/2}(z-1)^{\delta/2}w(z)$ (blue, dashed), computed numerically, and of the three-term power expansion near $z=1$ (solid, orange), obtained analytically using the connection coefficients computed at two instantons. The validity of the series expansion around $z=1$ (orange) is limited to a neighborhood of $z=1$, but going to higher orders in the expansion to extend the validity is straightforward. The values of the parameters are:
$a_1 = 0.5 + 1.24031i,\,a_2 = -0.5 + 1.55419i,\,E = 5.52396,\,m_3 = 0.92039 + 1.36765i,\, \Lambda=1.60238 + 1.25941i$.}}
\end{figure}
As a concluding remark, we notice that already the first instanton correction significantly improves the approximation.

\section{Applications to the black hole problem}\label{five}
There are several interesting physical quantities in the black hole problem which are governed by the Teukolsky equation. Having the explicit expression for the connection coefficients allows us to compute them exactly. We turn to this now.
\subsection{The greybody factor}
While all our analysis has been for classical black holes, it is known that quantum black holes emit thermal radiation from their horizons \cite{Hawking:1974sw}. However, the spacetime outside of the black hole acts as a potential barrier for the emitted particles, so that the emission spectrum as measured by an observer at infinity is no longer thermal, but is given by $\frac{\sigma(\omega)}{\exp{\frac{\omega-m\Omega}{T_H}}-1}$, where $\sigma(\omega)$ is the so-called greybody factor. Incidentally, it is the same as the absorption coefficient of the black hole, which tells us the ratio of a flux of particles incoming from infinity which penetrates the potential barrier and is absorbed by the black hole \cite{Hawking:1974sw} \cite{dong2015greybody}. More precisely, the radial equation with $s=0$ has a conserved flux, given by the "probability flux" when written as a Schr\"odinger equation: $\phi = \mathrm{Im}\psi^\dagger(z)\partial_z\psi(z)$ for $z$ on the real line. The absorption coefficient is then defined as the ratio between the flux $\phi_{abs}$ absorbed by the black hole (ingoing at the horizon) and the flux $\phi_{in}$ incoming from infinity. For non-zero spin, the potential (\ref{radial_potential}) becomes complex, and the flux is no longer conserved. In that case the absorption coefficient can be computed using energy fluxes \cite{Brito_2020}, but for simplicity we stick here to $s=0$.
\subsubsection{The exact result}
On physical grounds we impose the boundary condition that there is only an ingoing wave at the horizon:
\begin{equation}
    R(r\to r_+) \sim (r-r_+)^{-i\frac{\omega-m\Omega}{4\pi T_H}} \,,
\end{equation}
so the wavefunction near the horizon is given by
\begin{equation}
    \psi(z) = \hat{f}_{\alpha_{2+}}^{(t)} (z) = (z-1)^{\frac{1}{2}+a_2} \left(1+\mathcal{O}(z-1)\right) \,,
\end{equation}
with $a_2 = -i \frac{\omega - m \Omega}{4 \pi T_H}$ and recall that the time-dependent part goes like $e^{-i\omega t}$. This boundary condition is independent of whether $\omega - m \Omega$ is positive or negative: an observer near the horizon always sees an ingoing flux into the horizon, but when $\omega - m \Omega<0$ it is outgoing according to an observer at infinity. This phenomenon is known as superradiance \cite{Iyer:1986np}. In any case, this gives the flux
\begin{equation}
    \phi_{abs} = \mathrm{Im}a_2 
\end{equation}
ingoing at the horizon. Using our connection formula, we find that near infinity the wavefunction behaves as
\begin{equation}\label{psi_nearinfty}
    \begin{aligned}
    \psi(z) & = \frac{M_{\alpha_{2+}, \alpha_-} f_{\alpha_-}^{(u)} (z)}{\langle \Delta_\alpha, \Lambda_0, m_0 | V_{\alpha_{2+}} (1) | \Delta_{\alpha_1} \rangle} + \frac{ M_{\alpha_{2+}, \alpha_+}f_{\alpha_+}^{(u)} (z)}{\langle \Delta_\alpha, \Lambda_0, m_0 | V_{\alpha_{2+}} (1) | \Delta_{\alpha_1} \rangle} = \\
    & = M_{\alpha_{2+}, \alpha_-} \Lambda^{-\frac{1}{2} - a} \sum_{\pm} \mathcal{A}_{\alpha_-, m_{3 \pm}} e^{\pm \frac{\Lambda z}{2}} \left( \Lambda z \right)^{ \pm  m_3} \frac{\langle \Delta_{\alpha-}, \Lambda_0, m_{0\pm} | V_{\alpha_{2}} (1) | \Delta_{\alpha_1} \rangle}{\langle \Delta_\alpha, \Lambda_0, m_0 | V_{\alpha_{2+}} (1) | \Delta_{\alpha_1} \rangle} \left(1+\mathcal{O}(z^{-1})\right) + (\alpha \rightarrow -\alpha)\,.
    \end{aligned}
\end{equation}
At infinity, the ingoing part of the wave is easy to identify: recalling that $\Lambda = -2 i \omega (r_+ - r_-)$ it corresponds to the positive sign in the exponential. So the flux incoming from infinity is
\begin{equation}
\begin{aligned}
    \phi_{in} & = \mathrm{Im}\frac{\Lambda}{2} \left|  M_{\alpha_{2+}, \alpha_-} \mathcal{A}_{\alpha_-, m_{3+}} \Lambda^{-\frac{1}{2} - a + m_3} \frac{\langle \Delta_{\alpha-}, \Lambda_0, m_{0+} | V_{\alpha_{2}} (1) | \Delta_{\alpha_1} \rangle}{\langle \Delta_\alpha, \Lambda_0, m_0 | V_{\alpha_{2+}} (1) | \Delta_{\alpha_1} \rangle} + (\alpha \rightarrow -\alpha) \right|^2 = \\
    & = -\frac{1}{2} \left|\frac{\Gamma (1 + 2a)\Gamma (2a) \Gamma (1 + 2a_2)\Lambda^{ -a + m_3}}{\Gamma\left(\frac{1}{2} + m_3 + a\right)\prod_\pm \Gamma\left(\frac{1}{2} \pm a_1 + a_2 +a\right)} \frac{\langle \Delta_{\alpha-}, \Lambda_0, m_{0+} | V_{\alpha_{2}} (1) | \Delta_{\alpha_1} \rangle}{\langle \Delta_\alpha, \Lambda_0, m_0 | V_{\alpha_{2+}} (1) | \Delta_{\alpha_1} \rangle} + (a \rightarrow -a) \right|^2 \,.
\end{aligned}
\end{equation}
The minus sign comes from the fact that we have simplified $\Lambda$ and we have $\mathrm{Im}\Lambda = - |\Lambda|$. Note that also the flux at the horizon is negative (for non-superradiant modes). So the full absorption coefficient/greybody factor, defined as the flux going into the horizon normalized by the flux coming in from infinity is:
\begin{equation}\label{exact_sigma}
    \sigma = \frac{\phi_{abs}}{\phi_{in}} = \frac{\displaystyle{-\mathrm{Im}2a_2}}{\displaystyle{\left|\frac{\Gamma (1 +2a)\Gamma (2a) \Gamma (1 + 2a_2)\Lambda^{ -a + m_3}}{\Gamma\left(\frac{1}{2} + m_3 + a\right)\prod_\pm \Gamma\left(\frac{1}{2} \pm a_1 + a_2 +a\right)} \frac{\langle \Delta_{\alpha-}, \Lambda_0, m_{0+} | V_{\alpha_{2}} (1) | \Delta_{\alpha_1} \rangle}{\langle \Delta_{\alpha}, \Lambda_0, m_{0} | V_{\alpha_{2+}} (1) | \Delta_{\alpha_1} \rangle} + (a \rightarrow -a) \right|^2}}\,.
\end{equation}
This is the exact result, given as a power series in $\Lambda$. The correlators have to be understood as computed in the NS limit with $\epsilon_1=1$. The ratio of correlators can be written in terms of the NS free energy (see Appendix \ref{appendix_semiclassical}), and substituting the dictionary (\ref{radial_dictionary}) we get
\begin{equation}\label{exact_sigma_gravity}
\boxed{\begin{aligned}
    & \sigma = \frac{\phi_{abs}}{\phi_{in}} = \frac{\omega-m\Omega}{2\pi T_H}\times \\
    \times & \left|\frac{\Gamma (1+2a)\Gamma (2a) \Gamma (1-i\frac{\omega-m\Omega}{2\pi T_H}) (-2i\omega(r_+-r_-))^{-a -2iM\omega}e^{-i\omega(r_+-r_-)}\exp{ \left( \frac{\partial\mathcal{F}^{\mathrm{inst}}}{\partial a_1}\right)}|_{a_1=a,a_2=-a} }{\Gamma\left(\frac{1}{2} -2iM\omega + a\right)\Gamma\left(\frac{1}{2}-i\frac{\omega-m\Omega}{2\pi T_H} +2iM\omega +a\right)\Gamma\left(\frac{1}{2}-2iM\omega +a\right)} + (a \rightarrow -a) \right|^{-2}\,.
\end{aligned}}
\end{equation}
Here $\mathcal{F}^{\mathrm{inst}}(\Lambda,a_1,a_2,m_1,m_2,m_3)$ is the instanton part of the NS free energy as defined in Appendix \ref{AppendixNekrasov} computed for general $\Vec{a}=(a_1,a_2)$ and after taking the derivative one substitutes the values $\Vec{a}=(a,-a)$ appropriate for $SU(2)$. The same holds for the second summand but one substitutes $\Vec{a}=(-a,a)$ in the end. To write this result fully in terms of the parameters of the black hole problem using the dictionary (\ref{radial_dictionary}), one has to invert the relation $E=a^2-\Lambda \partial_{\Lambda} \mathcal{F}^{\mathrm{inst}}$ to obtain $a(E)$, which can be done order by order in $\Lambda$. In the literature, the absorption coefficient for Kerr black holes has been calculated using various approximations. As a consistency check, we show that our result reproduces the known results in the appropriate regimes.
\subsubsection{Comparison with asymptotic matching}
In \cite{Maldacena_1997}, the absorption coefficient is calculated via an asymptotic matching procedure. They work in a regime in which $\text{a}\omega \ll 1$ such that the angular eigenvalue $\lambda \approx \ell(\ell+1)$, and solve the Teukolsky equation for $s=0$ asymptotically in the regions near and far from the outer horizon. Then one also takes $M\omega \ll 1$ such that there exists an overlap between the far and near regions and one can match the asymptotic solutions. For us these limits imply that also $|\Lambda| = 4 \omega \sqrt{M^2-\text{a}^2} \ll 1$, so we expand our exact transmission coefficient to lowest order in $\text{a}\omega$, $M\omega$ and $\Lambda$. Since from the dictionary (\ref{radial_dictionary}) $E = a^2 + \mathcal{O}(\Lambda) = \frac{1}{4}+\ell(\ell+1)+\mathcal{O}(\text{a}\omega,M\omega)$, in this limit we have $a = \ell+\frac{1}{2}$. Then the second term in the denominator of (\ref{exact_sigma}) which contains $\Lambda^a$ vanishes for $\Lambda \rightarrow 0$ while the first one survives and passes to the numerator. The instanton part of the NS free energy also vanishes, $\mathcal{F}^{\mathrm{inst}}(\Lambda\to0)=0$. (\ref{exact_sigma_gravity}) then becomes
\begin{equation}
\begin{aligned}
    & \sigma \approx \frac{\omega-m\Omega}{2\pi T_H}(2\omega(r_+-r_-))^{2\ell+1} \left|\frac{\Gamma\left(\ell+1\right)\Gamma\left(\ell+1-i\frac{\omega-m\Omega}{2\pi T_H}\right)\Gamma\left(\ell+1\right)}{\Gamma (2\ell+2)\Gamma (2\ell+1) \Gamma (1-i\frac{\omega-m\Omega}{2\pi T_H}) }\right|^{2}\,.
\end{aligned}
\end{equation}
Using the relation $\frac{\Gamma(\ell+1)}{\Gamma( 2 \ell+2)} = \frac{\sqrt{\pi} }{2^{2 \ell+1}\Gamma(\ell+\frac{3}{2})}$ (and sending $i\to-i$ inside the modulus squared) we reduce precisely to the result of \cite{Maldacena_1997} (eq. 2.29):
\begin{equation}
    \boxed{\sigma \approx \frac{\omega - m \Omega}{2 T_H} \frac{ (r_+ - r_-)^{2\ell+1}\omega^{2\ell+1} }{2^{2\ell+1}}\left| \frac{\Gamma(\ell+1)\Gamma\left(\ell+1+i \frac{\omega - m \Omega}{2 \pi T_H}\right)}{\Gamma\left(\ell+\frac{3}{2}\right)\Gamma(2\ell+1) \Gamma \left(1 + i \frac{\omega - m \Omega}{2 \pi T_H}\right)} \right|^2 \,,} 
\end{equation}
which is valid for $M\omega,\text{a}\omega \ll 1$.

\subsubsection{Comparison with semiclassics}
We now show that the exact absorption coefficient reduces to the semiclassical result obtained via a standard WKB analysis of the equation
\begin{equation}
    \epsilon_1^2 \partial_z^2\psi(z)+V(z)\psi(z) = 0\,.
\end{equation}
where we have reintroduced the small parameter $\epsilon_1$ which plays the role of the Planck constant to keep track of the orders in the expansion. For the Teukolsky equation (which has $\epsilon_1=1$) the semiclassical regime is the regime in which $\ell \gg 1$. Following \cite{dumlu2020stokes}, we also take $M\omega \ll 1$ and $s=0$ such that there are two zeroes of the potential between the outer horizon and infinity for real values of $z$ which we denote by $z_1$ and $z_2$ with $z_2 > z_1$, between which there is a potential barrier for the particle ($V(z)$ becomes negative, notice the "wrong sign" in front of the second derivative). Without these extra conditions, the potential generically becomes complex, or does not form a barrier. The main difference with the regime used for the asymptotic matching procedure in the previous section is that there we worked to leading order in $M\omega,a\omega$. Now we still assume them to be small but keep all orders, while working to first subleading order in $\epsilon_1$.
\begin{figure}[H]
\centering
\includegraphics[width=\textwidth]{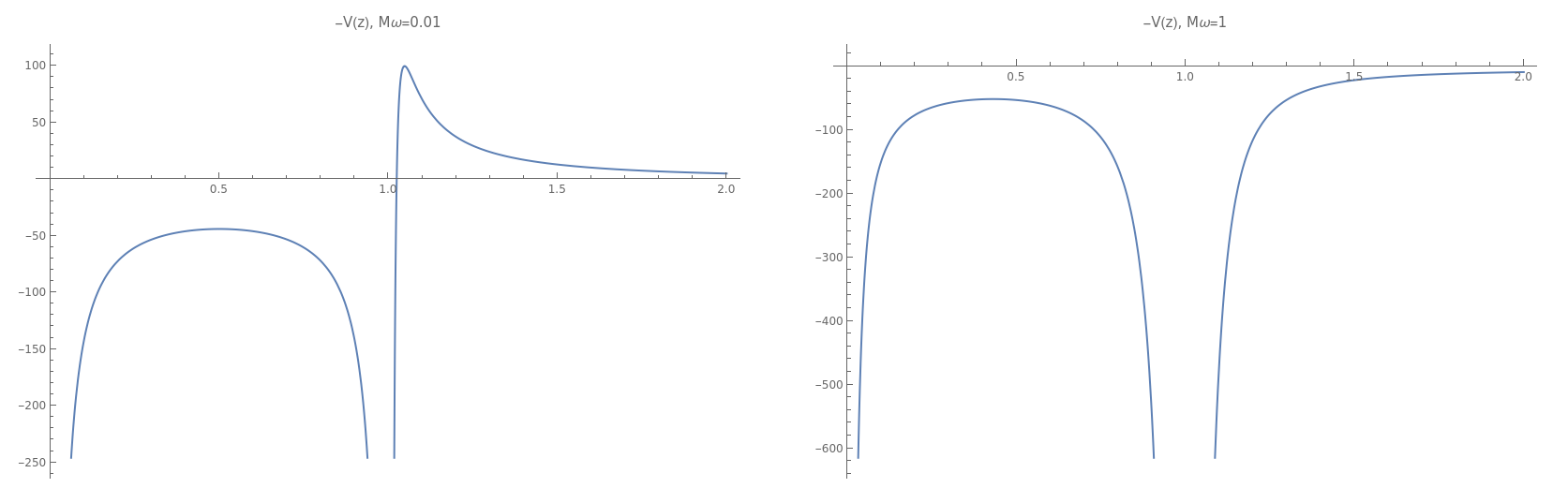}
\caption{{\small Forms of the potential $-V(z)$ for $M=1,\,a=0.5,\, \lambda=10,\,m=0\,,s=0$, and $\omega=0.01$ (left) and $\omega=1$ (right). We see that for $M\omega$ not small enough, the potential does not form a barrier.}}
\end{figure}
The standard WKB solutions are
\begin{equation}
    \psi(z) \propto V(z)^{-\frac{1}{4}}\exp \left( \pm \frac{i}{\epsilon_1}\int_{z_*}^z \sqrt{V(z')}dz' \right) \,,
\end{equation}
where $z_*$ is some arbitrary reference point, usually taken to be a turning point of the potential, here corresponding to a zero. The absorption coefficient is given by the transmission coefficient from infinity to the horizon and captures the tunneling amplitude through this potential barrier. It is simply given by
\begin{equation}
    \sigma \approx \exp \left( \frac{2i}{\epsilon_1} \int_{z_1}^{z_2} \sqrt{V(z')}dz' \right) =\exp \left(-\frac{2}{\epsilon_1} \int_{z_1}^{z_2} \sqrt{|V(z')|}dz'\right) \,.
\end{equation}
On the other hand it is known that in the semiclassical limit the potential of the BPZ equation reduces to the Seiberg-Witten differential of the AGT dual gauge theory \cite{Alday_2010}, which for us is $SU(2)$ gauge theory with $N_f=3$: $V(z) \rightarrow - \phi^2_{SW}(z)$. The integral between the two zeroes then corresponds to half a B-cycle, so we identify
\begin{equation}
    \boxed{\sigma \approx \exp \left( -\frac{2}{\epsilon_1} \int_{z_1}^{z_2} \phi_{SW}(z')dz'\right) = \exp \left( -\frac{1}{\epsilon_1} \oint_B \phi_{SW}(z')dz' \right) =: \exp \left( -\frac{a_D}{\epsilon_1} \right) \,,}
\end{equation}
where we have chosen an orientation of the B-cycle. Our exact absorption coefficient reduces to this expression in the semiclassical limit $\epsilon_1 \rightarrow 0$. The detailed calculation is deferred to Appendix \ref{appendix_semiclassical}.

\subsection{Quantization of quasinormal modes}
With the explicit expression of the connection matrix (\ref{eq:irregconnectionmatrix}) in our hands we can extract the quantization condition for the quasinormal modes. The correct boundary conditions for quasinormal modes is only an ingoing wave at the horizon and only an outgoing one at infinity (see e.g. \cite{Berti_2009}, eq. (80)), that is
\begin{equation}
    \begin{aligned}
    &R_{\mathrm{QNM}}(r\to r_+) \sim (r-r_+)^{-i\frac{\omega-m\Omega}{4\pi T_H}-s}\\
    &R_{\mathrm{QNM}}(r\to \infty) \sim r^{-1-2s+2iM\omega} e^{i\omega r}\,.
    \end{aligned}
\end{equation}
In terms of the function $\psi(z)$ satisfying the Teukolsky equation in Schr\"odinger form:
\begin{equation}
\begin{aligned}
&\psi_{\mathrm{QNM}} (z \to 1) \sim (z-1)^{\frac{1}{2}+a_2} \,, \\
&\psi_{\mathrm{QNM}} (z \to \infty) \sim e^{- \Lambda z /2} \left( \Lambda z \right)^{-m_3} \,.
\end{aligned}
\label{eq:QNMboundary}
\end{equation}
However, imposing the ingoing boundary condition at the horizon and using the connection formula, we get that near infinity
\begin{equation}
\begin{aligned}
& \psi_{\mathrm{QNM}} (z \to \infty) \sim \\
\sim & \bigg( \Lambda^{a} M_{\alpha_{2+}, \alpha_+} \mathcal{A}_{\alpha_+ m_{0 -}}  \frac{\langle \Delta_{\alpha+}, \Lambda_0, m_{0-} | V_{\alpha_{2}} (1) | \Delta_{\alpha_1} \rangle}{\langle \Delta_{\alpha}, \Lambda_0, m_{0} | V_{\alpha_{2+}} (1) | \Delta_{\alpha_1} \rangle}+ \Lambda^{- a} M_{\alpha_{2+}, \alpha_-} \mathcal{A}_{\alpha_- m_{0 -}} \frac{\langle \Delta_{\alpha-}, \Lambda_0, m_{0-} | V_{\alpha_{2}} (1) | \Delta_{\alpha_1} \rangle}{\langle \Delta_{\alpha}, \Lambda_0, m_{0} | V_{\alpha_{2+}} (1) | \Delta_{\alpha_1} \rangle} \bigg)\times\\&\times e^{- \Lambda z /2} \left( \Lambda z \right)^{-m_3} + \\ +&  \bigg( \Lambda^{a} M_{\alpha_{2+}, \alpha_+} \mathcal{A}_{\alpha_+ m_{0 +}}  \frac{\langle \Delta_{\alpha+}, \Lambda_0, m_{0+} | V_{\alpha_{2}} (1) | \Delta_{\alpha_1} \rangle}{\langle \Delta_{\alpha}, \Lambda_0, m_{0} | V_{\alpha_{2+}} (1) | \Delta_{\alpha_1} \rangle} + \Lambda^{- a} M_{\alpha_{2+}, \alpha_-} \mathcal{A}_{\alpha_- m_{0 +}} \frac{\langle \Delta_{\alpha-}, \Lambda_0, m_{0+} | V_{\alpha_{2}} (1) | \Delta_{\alpha_1} \rangle}{\langle \Delta_{\alpha}, \Lambda_0, m_{0} | V_{\alpha_{2+}} (1) | \Delta_{\alpha_1} \rangle} \bigg)\times\\&\times e^{\Lambda z /2} \left( \Lambda z \right)^{m_3} \,,
\end{aligned}
\end{equation}
which contains both an ingoing an an outgoing wave at infinity. In order to impose the correct boundary condition (\ref{eq:QNMboundary}) we need to impose that the coefficient of the ingoing wave vanishes:
\begin{equation}
\begin{aligned}
&\Lambda^{a} M_{\alpha_{2+}, \alpha_+} \mathcal{A}_{\alpha_+ m_{0 +}}  \frac{\langle \Delta_{\alpha+}, \Lambda_0, m_{0+} | V_{\alpha_{2}} (1) | \Delta_{\alpha_1} \rangle}{\langle \Delta_{\alpha}, \Lambda_0, m_{0} | V_{\alpha_{2+}} (1) | \Delta_{\alpha_1} \rangle} + \Lambda^{- a} M_{\alpha_{2+}, \alpha_-} \mathcal{A}_{\alpha_- m_{0 +}} \frac{\langle \Delta_{\alpha-}, \Lambda_0, m_{0+} | V_{\alpha_{2}} (1) | \Delta_{\alpha_1} \rangle}{\langle \Delta_{\alpha}, \Lambda_0, m_{0} | V_{\alpha_{2+}} (1) | \Delta_{\alpha_1} \rangle}=0 \\
&\Longrightarrow 1 + \Lambda^{-2a} \frac{M_{\alpha_{2+}, \alpha_-} \mathcal{A}_{\alpha_- m_{0 +}} \langle \Delta_{\alpha-}, \Lambda_0, m_{0+} | V_{\alpha_{2}} (1) | \Delta_{\alpha_1} \rangle}{M_{\alpha_{2+}, \alpha_+} \mathcal{A}_{\alpha_+ m_{0 +}}  \langle \Delta_{\alpha+}, \Lambda_0, m_{0+} | V_{\alpha_{2}} (1) | \Delta_{\alpha_1} \rangle} = 0 \,.
\end{aligned}
\label{eq:quantizcft}
\end{equation} 
Identifying in the NS limit
\begin{equation}
\begin{aligned}
& \frac{\langle \Delta_{\alpha-}, \Lambda_0, m_{0+} | V_{\alpha_{2}} (1) | \Delta_{\alpha_1} \rangle}{\langle \Delta_{\alpha+}, \Lambda_0, m_{0+} | V_{\alpha_{2}} (1) | \Delta_{\alpha_1} \rangle} = \frac{\mathcal{Z} (\Lambda, a+\frac{\epsilon_2}{2}, m_1, m_2,  m_3+\frac{\epsilon_2}{2})}{\mathcal{Z} (\Lambda, a-\frac{\epsilon_2}{2}, m_1, m_2,  m_3+\frac{\epsilon_2}{2})} =\\
= & \exp \frac{1}{\epsilon_1 \epsilon_2}\left(\mathcal{F}^{\mathrm{inst}} (\Lambda, a + \frac{\epsilon_2}{2}, m_1, m_2, m_3+\frac{\epsilon_2}{2}) - \mathcal{F}^{\mathrm{inst}} (\Lambda, a - \frac{\epsilon_2}{2}, m_1, m_2, m_3+\frac{\epsilon_2}{2}) \right) \to \\ \to &\exp \frac{\partial_a \mathcal{F}^{\mathrm{inst}}(\Lambda, a, m_1, m_2, m_3)}{\epsilon_1} \,.
\label{eq:prepinstqnms}
\end{aligned}
\end{equation}
Moreover,
\begin{equation}
\begin{aligned}
\frac{M_{\alpha_{2+}, \alpha_-} \mathcal{A}_{\alpha_- m_{0 +}}}{M_{\alpha_{2+}, \alpha_+} \mathcal{A}_{\alpha_+ m_{0 +}}} &= \frac{\Gamma \left( \frac{2a}{\epsilon_1} \right) \Gamma \left(1 + \frac{2a}{\epsilon_1} \right)}{\Gamma \left(- \frac{2a}{\epsilon_1} \right) \Gamma \left(1 - \frac{2a}{\epsilon_1} \right)} \frac{\Gamma \left( \frac{1}{2} + \frac{a_2 + a_1 - a}{\epsilon_1} \right)\Gamma \left( \frac{1}{2} + \frac{a_2 - a_1 - a}{\epsilon_1} \right)}{\Gamma \left( \frac{1}{2} + \frac{a_2 + a_1 + a}{\epsilon_1} \right)\Gamma \left( \frac{1}{2} + \frac{a_2 - a_1 + a}{\epsilon_1} \right)} \frac{\Gamma \left( \frac{1}{2} + \frac{m_3 - a}{\epsilon_1} \right)}{\Gamma \left( \frac{1}{2} + \frac{m_3 + a}{\epsilon_1} \right)} = \\ &= \frac{\Gamma \left( \frac{2a}{\epsilon_1} \right) \Gamma \left(1 + \frac{2a}{\epsilon_1} \right)}{\Gamma \left(- \frac{2a}{\epsilon_1} \right) \Gamma \left(1 - \frac{2a}{\epsilon_1} \right)} \prod_{i=1}^3 \frac{\Gamma \left( \frac{1}{2} + \frac{m_i - a}{\epsilon_1} \right)}{\Gamma \left( \frac{1}{2} + \frac{m_i + a}{\epsilon_1} \right)} = e^{-i \pi} \left( \frac{\Gamma \left(1 + \frac{2a}{\epsilon_1} \right) }{\Gamma \left(1 - \frac{2a}{\epsilon_1} \right)} \right)^2 \prod_{i=1}^3 \frac{\Gamma \left( \frac{1}{2} + \frac{m_i - a}{\epsilon_1} \right)}{\Gamma \left( \frac{1}{2} + \frac{m_i + a}{\epsilon_1} \right)} = \\ &= \exp \left[- i \pi + 2 \log \frac{\Gamma \left(1 + \frac{2a}{\epsilon_1} \right) }{\Gamma \left(1 - \frac{2a}{\epsilon_1} \right)}  + \sum_{i=1}^3 \log  \frac{\Gamma \left( \frac{1}{2} + \frac{m_i - a}{\epsilon_1} \right)}{\Gamma \left( \frac{1}{2} + \frac{m_i + a}{\epsilon_1} \right)}  \right]\,.
\end{aligned}
\label{eq:ratio1coefficients}
\end{equation}
Including also the $\Lambda$ factor (restoring the factor of $\epsilon_1$), we identify the exponent with (see Appendix \ref{AppendixNekrasov})
\begin{equation}
\frac{1}{\epsilon_1} \left[ -i \pi \epsilon_1 -  2 a \log \frac{\Lambda}{\epsilon_1}  + 2 \epsilon_1 \log \frac{\Gamma \left(1 + \frac{2a}{\epsilon_1} \right) }{\Gamma \left(1 - \frac{2a}{\epsilon_1} \right)} + \epsilon_1 \sum_{i=1}^3 \log  \frac{\Gamma \left( \frac{1}{2} + \frac{m_i - a}{\epsilon_1} \right)}{\Gamma \left( \frac{1}{2} + \frac{m_i + a}{\epsilon_1} \right)}\right] = -i \pi + \frac{1}{\epsilon_1}\partial_a \mathcal{F}^{\mathrm{1-loop}} \,.
\label{eq:tobematchedGrassi}
\end{equation}
The instanton and one loop part combine to give the full NS free energy, and hence (\ref{eq:quantizcft}) can be conveniently rewritten for $\epsilon_1 = 1$ (as required by the dictionary), as
\begin{equation}
1 - e^{\partial_a \mathcal{F}} = 0 \Rightarrow \partial_a \mathcal{F} = 2 \pi i n \,, n \in \mathbb{Z}\,.
\end{equation}
To solve for the quasinormal mode frequencies, we need to invert the relation $E=a^2-\Lambda \partial_\Lambda \mathcal{F}^{\mathrm{inst}}$ to obtain $a(E)$. Then the quantization condition for the quasinormal mode frequencies that we have derived reads
\begin{equation}
    \boxed{\partial_a \mathcal{F}\left(-2i\omega(r_+-r_-),a(E),-i \frac{\omega - m \Omega}{2 \pi T_H} + 2i M \omega,-2iM\omega-s,-2iM\omega+s,1\right)=2\pi i n\,, n \in \mathbb{Z}\,,}
\end{equation}
with $E=\frac{1}{4} + \lambda + s(s+1) + \text{a}^2 \omega^2 - 8M^2 \omega^2 - \left(2M\omega^2 + i s \omega \right) (r_+-r_-)$.
This gives an equation that is solved for a discrete set of $\omega_n$, in agreement with \cite{aminov2020black}\footnote{In order to match with \cite{aminov2020black}, it is important to notice that they use the variable $-ia$ instead of $a$, have a different $U(1)$ factor as previously noticed, and a sign difference in the definition of the free energy $\mathcal{F}$. Moreover, their $\partial_a \mathcal{F}$ is shifted by a factor of $- i \pi$ with respect to ours.}.

\subsection{Angular quantization}\label{angular_quantization}
Yet another application of the connection formulae is the computation of the angular eigenvalue $\lambda$. To this end, we impose regularity of the angular eigenfunctions at $z = 0, 1$. According to the angular dictionary (\ref{eq:angulardictionaryy}), 
\begin{equation}
\frac{1 \pm 2 a_1}{2} = \frac{1}{2} \mp \frac{m-s}{2} \,, \, \frac{1 \pm 2 a_2}{2} = \frac{1}{2} \mp \frac{m+s}{2} \,,
\end{equation}
therefore, according to (\ref{eq:angchangeofvar}) the behavior of $S_\lambda$ as $z \to 0$ is given by
\begin{equation}
S_\lambda (z \to 0) \propto z^{\mp \frac{m-s}{2}} \,.
\end{equation}
Since $\lambda_{s,m} = \lambda^*_{s,-m}$, $\lambda_{-s,m} = \lambda_{s,m} + 2 s$ \cite{Berti:2005gp}, we can restrict without loss of generality to the case $m, -s \ge 0$. Regularity of $S_\lambda$ as $z \to 0$ requires the boundary condition
\begin{equation}
y_{m>s}(z \to 0) = \hat{f}_{\alpha_{1-}}^{(s)} (z) \simeq z^{\frac{1}{2} + \frac{m-s}{2}}\,.
\end{equation}
Therefore near $z \to 1$, 
\begin{equation}
\begin{aligned}
&y_{m>s}(z \to 1) = \\ &= N_{a_{1-}, a_{2-}} \frac{\langle \Delta_\alpha, \Lambda_0, m_0 | V_{\alpha_{2-}} (1) | \Delta_{\alpha_1} \rangle}{\langle \Delta_\alpha, \Lambda_0, m_0 | V_{\alpha_{2}} (1) | \Delta_{\alpha_{1-}} \rangle} \hat{f}_{\alpha_{2-}}^{(t)} (z) + N_{a_{1-}, a_{2+}} \frac{\langle \Delta_\alpha, \Lambda_0, m_0 | V_{\alpha_{2+}} (1) | \Delta_{\alpha_1} \rangle}{\langle \Delta_\alpha, \Lambda_0, m_0 | V_{\alpha_{2}} (1) | \Delta_{\alpha_{1-}} \rangle}  \hat{f}_{\alpha_{2+}}^{(t)} (z) \simeq \\ &\simeq \frac{\Gamma (-m-s) \Gamma (1+m-s)}{\Gamma (\frac{1}{2} -a -s) \Gamma (\frac{1}{2} +a -s)} \frac{\langle \Delta_\alpha, \Lambda_0, m_0 | V_{\alpha_{2-}} (1) | \Delta_{\alpha_1} \rangle}{\langle \Delta_\alpha, \Lambda_0, m_0 | V_{\alpha_{2}} (1) | \Delta_{\alpha_{1-}} \rangle} (1-z)^{\frac{1}{2} + \frac{m+s}{2}} + \\ &+ \frac{\Gamma(m+s) \Gamma(1+m-s)}{\Gamma(\frac{1}{2} - a + m) \Gamma (\frac{1}{2} + a + m)} \frac{\langle \Delta_\alpha, \Lambda_0, m_0 | V_{\alpha_{2+}} (1) | \Delta_{\alpha_1} \rangle}{\langle \Delta_\alpha, \Lambda_0, m_0 | V_{\alpha_{2}} (1) | \Delta_{\alpha_{1-}} \rangle} (1-z)^{\frac{1}{2} - \frac{m+s}{2}} \,.
\end{aligned}
\label{eq:90}
\end{equation}
Let us start by assuming $m+s>0$. Then the second term in (\ref{eq:90}) has a pole at $z =1$ for generic values of $a$, and the first gamma function is divergent as it stands. However both divergences are cured by imposing that
\begin{equation}
a = \ell + \frac{1}{2} \,,
\end{equation}
for some positive integer $\ell \ge m \ge -s$. Analogously if $m+s \le 0$, regularity is ensured by imposing $a = \ell + \frac{1}{2}$ with $\ell \ge m \ge -s$. Therefore in general the quantization condition for the angular eigenvalue is
\begin{equation}
a(\Lambda,E,m_1,m_2,m_3) = \ell + \frac{1}{2} \,, \, \ell \ge \text{max} (m \,, -s) \,.
\end{equation}
As before, $a$ is obtained by inverting the expression $E=a^2-\Lambda \partial_\Lambda \mathcal{F}^{\mathrm{inst}}$ order by order in $\Lambda$. Let us denote by
\begin{equation}
\lambda_0 = \lambda (\Lambda = 0) = \ell (\ell + 1) - s(s+1) \,.
\end{equation}
Then the above quantization condition for the angular eigenvalue $\lambda$ can be more conveniently written as
\begin{equation}
\lambda - \lambda_0 = 2cs-c^2-\Lambda\partial_\Lambda \mathcal{F}^{\mathrm{inst}}\left(\Lambda,\ell+\frac{1}{2},-m,-s,-s\right)\bigg|_{\Lambda=4c} \,,
\end{equation}
which is the result already obtained in \cite{aminov2020black}.

\subsection{Love numbers}
Applying an external gravitational field to a self-gravitating body generically causes it to deform, much in the same way as an external electric field polarizes a dielectric material. The response of the body to the external gravitational tidal field is captured by the so-called tidal response coefficients or Love numbers, named after A. E. H. Love who first studied them in the context of the Earth's response to the tides \cite{Love_1909}. In general relativity, the tidal response coefficients are generally complex, and the real part captures the conservative response of the body, whereas the imaginary part captures dissipative effects. There is some naming ambiguity where sometimes only the real, conservative part is called the Love number, whereas sometimes the full complex response coefficient is called Love number. For us the Love number will be the full complex response coefficient. For four-dimensional Kerr black holes, the conservative (real part) of the response coefficient to static external perturbations has been found to vanish \cite{Le_Tiec_2021, charalambous2021vanishing}. Moreover, Love numbers are measurable quantities that can be probed with gravitational wave observations \cite{Flanagan_2008, Cardoso_2017}. Using our conformal field theory approach to the Teukolsky equation we compute the Love number of a slowly rotating Kerr black hole at linear order in the frequency of the perturbation. The extension of our computation to higher orders is straightforward. 
\subsubsection{Definition of Love number and the intermediate region}
For the definition of Love numbers we follow \cite{charalambous2021vanishing} and \cite{Le_Tiec_2021}, to which we refer for a more complete introduction. In the case of a static external perturbation ($\omega=0$), one imposes the ingoing boundary condition on the radial part of the perturbing field at the horizon, which then behaves near infinity as
\begin{equation}
\begin{aligned}\label{staticR}
    R(r\to\infty) &= A r^{\ell-s}(1+\mathcal{O}(r^{-1})) + B r^{-\ell-s-1}(1+\mathcal{O}(r^{-1})) \\
    &= A r^{\ell-s}\left[(1+\mathcal{O}(r^{-1})) + k_{\ell m}^{(s)} \left(\frac{r}{r_+-r_-}\right)^{-2\ell-1}(1+\mathcal{O}(r^{-1}))\right]
\end{aligned}
\end{equation}
for some constants $A$ and $B$. The Love number $k_{\ell m}^{(s)}$ is then defined as the coefficient of $(r/(r_+-r_-))^{-2\ell-1}$ (note that this differs from the definition in \cite{charalambous2021vanishing} where they define it as the coefficient of $(r/2M)^{-2\ell-1}$ instead). In the non-static case however, the definition of Love number is less clear, since the behaviour of the radial function at infinity is now qualitatively different from (\ref{staticR}): it is oscillatory (cf. (\ref{psi_nearinfty})) due to the term $\propto \omega^2$ in the potential (\ref{radial_As}). For small frequencies we can however define an intermediate regime $r \gg M$, $r\omega \ll 1$ in which the multipole expansion (\ref{staticR}) is still valid and we can read off the Love numbers in the same way as in the static case. Recall the Teukolsky equation written as a Schr\"odinger equation:
\begin{equation}
    \frac{d^2 \psi(z)}{dz^2} + V_{CFT}(z) \psi(z) = 0
\end{equation}
with the potential \eqref{eq:schroedingerexp}
\begin{equation}
    V_{CFT}(z) = - \frac{1}{z} \frac{1}{z-1} \big(- \Lambda \partial_{\Lambda} \mathcal{F}^{\mathrm{inst}} + \hat{\Delta}_2 + \hat{\Delta}_1 -\hat{\Delta} \big) + \frac{\hat{\Delta}_2}{(z-1)^2} + \frac{\hat{\Delta}_1}{z^2} - \frac{m_3 \Lambda}{z} - \frac{\Lambda^2}{4} \,.
\end{equation}
The intermediate regime corresponds to $z \gg 1$, $\Lambda z \ll 1$. Expanding in these variables the potential reads:
\begin{equation}
    \frac{V_{CFT}(z)}{\Lambda^2} = \frac{\frac{1}{4}-E}{\Lambda^2 z^2} \left( 1+\mathcal{O}(z^{-1},\Lambda z) \right)\,.
\end{equation}
We see that in this regime the leading term in the potential is the one $\propto 1/z^2$, and the multipole expansion holds. In a sense we are taking $z$ to be big enough to be far from the horizon, but not so far as to reach the oscillatory region at infinity, as already mentioned in \cite{chia2020tidal}. In the static case this intermediate region where the multipole expansion is valid extends all the way to infinity. On the CFT side, the conformal blocks in this regime are computed by expanding the irregular state as in (\ref{irregular_state}) and doing the OPE of the degenerate state near infinity term by term. This gives an expansion in $\Lambda z$ and $z^{-1}$.
\subsection{Slowly rotating Kerr Love numbers}
Let us compute the Kerr Love numbers up to first order in $M \omega \sim M \Omega$. In order to do this we have to consider only the first instanton correction since $\Lambda \propto M \omega$. The wavefunction up to one instanton can be derived from the conformal blocks in the intermediate regime. Schematically, 
\begin{equation}
    \psi (z) \sim \frac{\langle \Delta ,\Lambda_0, m_0 | \phi(z) V_2(1) | \Delta_1 \rangle}{\langle \Delta ,\Lambda_0 ,m_0 | V_2(1) | \Delta_1 \rangle} \simeq \frac{ \left( \langle \Delta| + \frac{m_0 \Lambda_0}{2 \Delta} \langle \Delta| L_1 \right) \phi(z) V_2(1) | \Delta_1 \rangle}{\left( \langle \Delta| + \frac{m_0 \Lambda_0}{2 \Delta} \langle \Delta| L_1 \right) V_2(1) | \Delta_1 \rangle} \,.
\end{equation}
Imposing the ingoing boundary condition at the horizon, this gives the following wavefunction in the intermediate regime:
\begin{equation}
\begin{aligned}
    \psi (z) = & \left[ 1 + \frac{m_3 \Lambda}{\frac{1}{2} - 2 a^2} \left( \left(1-\frac{1}{z} \right) \partial_{1/z} + z - \frac{1}{2} \right) \right] \sum_{\theta = \pm} M_{a_{2+} a_\theta}  z^{\frac{1}{2}-\theta a} \left(1-\frac{1}{z} \right)^{\frac{1}{2}+a_2} \times \\ &\times {}_2 F_1 \left( \frac{1}{2}+a_2+\theta a -a_1, \frac{1}{2}+a_2+\theta a +a_1; 1+2 \theta a; \frac{1}{z} \right) + \mathcal{O} \left( \Lambda^2 \right) \,.
\end{aligned}
\end{equation}
Note that the first instanton contributes at this order only if $s \ne 0$ since for zero spin $m_3 \Lambda \sim \mathcal{O} (M^2 \omega^2)$. For a slowly rotating black hole the connection coefficients start with $\mathcal{O}((M \omega)^0) = \mathcal{O}((M \Omega)^0)$ terms. Indeed substituting the dictionary we find
\begin{equation}
\begin{aligned}
    M_{a_{2 +} a_+} &= \frac{\Gamma(-1-2 \ell -2\Delta \ell) \Gamma(1 - 2 i \frac{\omega - m \Omega}{4 \pi T_H} - s)}{\Gamma(-\ell-\Delta \ell - 2i \frac{\omega - m \Omega}{4 \pi T_H} + 2 i M \omega) \Gamma(-\ell - \Delta \ell - 2 i M \omega - s)} = \\ &= \frac{\ell ! (\ell + s)!}{(2 \ell + 1)!} (-1)^{s+1} \frac{(2 i M \omega)(- 2i \frac{\omega - m \Omega}{4 \pi T_H} + 2 i M \omega)}{2 \Delta \ell} + \mathcal{O}(M \omega) \,, \\ 
    M_{a_{2 +} a_-} &= \frac{\Gamma(1+2 \ell) \Gamma(1-s)}{\Gamma(\ell + 1) \Gamma(\ell-s +1)} +\mathcal{O}(M \omega) \,,
\end{aligned}
\end{equation}
where $a = \ell + 1/2 + \Delta \ell$. It turns out that the first correction to $a$ vanishes, so $\Delta \ell \sim \mathcal{O} (M^2 \omega^2)$. Also note that all the Gamma functions are finite since $s \le 0$. Plugging in the dictionary and expanding the hypergeometrics gives
\begin{equation}
\begin{aligned}
    &{}_2 F_1 \left( \frac{1}{2}+a_2+a -a_1, \frac{1}{2}+a_2+ a +a_1; 1+2 a; \frac{1}{z} \right) \simeq {}_2 F_1 \left( 1+\ell-s-2 i M \omega, 1+\ell - 2i \frac{\omega - m \Omega}{4 \pi T_H} + 2 i M \omega; 2 + 2 \ell; \frac{1}{z} \right) \,, \\
    &{}_2 F_1 \left( \frac{1}{2}+a_2-a -a_1, \frac{1}{2}+a_2 - a +a_1; 1-2 a; \frac{1}{z} \right) \simeq \sum_{n=0}^{2 \ell} \frac{(-\ell -s -2 i M \omega)_{(n)} (-\ell - 2i \frac{\omega - m \Omega}{4 \pi T_H} + 2 i M \omega)_{(n)}}{(-2\ell)_{(n)}} \frac{z^{-n}}{n!} + \\ &+ \frac{\Gamma(-2 \ell - 2 \Delta \ell) \Gamma(1+\ell-s-2iM \omega) \Gamma(1+\ell- 2i \frac{\omega - m \Omega}{4 \pi T_H} +2iM \omega)}{\Gamma(-\ell-s-2iM \omega) \Gamma(-\ell - 2i \frac{\omega - m \Omega}{4 \pi T_H} +2iM \omega) \Gamma(2 \ell+2)} z^{-2 \ell-1} \times \\ &\times {}_2 F_1 \left(  1+\ell-s-2 i M \omega, 1+\ell - 2i \frac{\omega - m \Omega}{4 \pi T_H} + 2 i M \omega; 2 + 2 \ell; \frac{1}{z}  \right) \,.
\end{aligned}
\end{equation}
Note that 
\begin{equation}
    \frac{\Gamma(-2 \ell - 2 \Delta \ell) \Gamma(1+\ell-s-2iM \omega) \Gamma(1+\ell- 2i \frac{\omega - m \Omega}{4 \pi T_H} +2iM \omega)}{\Gamma(-\ell-s-2iM \omega) \Gamma(-\ell - 2i \frac{\omega - m \Omega}{4 \pi T_H} +2iM \omega) \Gamma(2 \ell+2)} M_{a_{2+} a_-} = - M_{a_{2+} a_+} + \mathcal{O} \left( (M \omega)^2 \right) \,,
\end{equation}
therefore at this order the hypergeometrics simplify one against the other up to a finite polynomial, hence
\begin{equation}
\begin{aligned}
    &\psi(z) = \left[ 1 + \frac{m_3 \Lambda}{\frac{1}{2} - 2 a^2} \left( \left(1-\frac{1}{z} \right) \partial_{1/z} + z - \frac{1}{2} \right) \right] \times \\& \times M_{a_{2+} a_-} z^{\frac{1}{2}+a} \left(1-\frac{1}{z} \right)^{\frac{1}{2}+a_2} \sum_{n=0}^{2 \ell} \frac{(-\ell -s -2 i M \omega)_{(n)} (-\ell - 2i \frac{\omega - m \Omega}{4 \pi T_H} + 2 i M \omega)_{(n)}}{(-2\ell)_{(n)}} \frac{z^{-n}}{n!} + \mathcal{O} (M^2 \omega^2) \,.
\end{aligned}
\label{eq:wvln1storder}
\end{equation}
The radial wavefunction is given by
\begin{equation}\label{eq:changeoffunction}
    R(r) = \Delta^{-\frac{s+1}{2}} (r) \psi(z) \,,
\end{equation}
where
\begin{equation}
    z = \frac{r}{2M} + \mathcal{O} \left( M^2 \Omega^2 \right)  \,, \, \, \Delta(r)^{-\frac{s+1}{2}} = (r_+ - r_-)^{-s-1} z^{-s-1} \left( 1-\frac{1}{z} \right)^{-\frac{s+1}{2}} \,.
\end{equation}
To find the Love numbers, we need the ratio between the coefficient of $r^{-\ell-s-1}$ (the response) and the coefficient of $r^{\ell-s}$ (the source). The term coming from the first instanton in (\ref{eq:wvln1storder}) will not contribute at this order. Indeed this term gives
\begin{equation}
\begin{aligned}
     &\psi(z) \supset \frac{-4 i M^2 \omega s}{\ell (\ell+1)} \left( \left(1-\frac{1}{z} \right) \partial_{1/z} + z - \frac{1}{2} \right) M_{a_{2+} a_-} z^{\ell+1} \left(1-\frac{1}{z} \right)^{\frac{1-s}{2}} \sum_{n=0}^{\ell+s} \frac{(-\ell -s)_{(n)} (-\ell)_{(n)}}{(-2\ell)_{(n)}} \frac{z^{-n}}{n!} + \mathcal{O} (M^2 \omega^2)= \\ &= \frac{-4 i M^2 \omega s}{\ell (\ell+1)} M_{a_{2+} a_-} z^{\ell+1} \left(1-\frac{1}{z} \right)^{\frac{1-s}{2}} \left( -z \ell + \frac{2 \ell + s}{2} + \left(1- \frac{1}{z}\right) \partial_{1/z} \right) \sum_{n=0}^{\ell+s} \frac{(-\ell -s)_{(n)} (-\ell)_{(n)}}{(-2\ell)_{(n)}} \frac{z^{-n}}{n!} + \mathcal{O} (M^2 \omega^2) \,.
\end{aligned}
\end{equation}
After taking into account the factor of $\Delta$ from (\ref{eq:changeoffunction}), one sees that this contribution to $R(r)$ does not contain the power that we are interested in. Focusing on the zero instanton contribution, the $(1-1/z)$ prefactor has an $\mathcal{O} (M \omega)$ term in the exponent that has to be expanded, resulting in
\begin{equation}
\begin{aligned}
    &R(r) \supset  i \frac{\omega-m\Omega}{4 \pi T_H} \frac{M_{a_{2+} a_-}}{(r_+ - r_-)^{s+1}} \frac{r^{\ell-s}}{((2M)^{\ell+1}} \left(1 + s \frac{2M}{r} + \frac{s(s+1)}{2} \left( \frac{2M}{r}\right)^2 \right) \sum_{k=1}^\infty \sum_{n=0}^{2 \ell} \frac{(-\ell -s)_{(n)} (-\ell )_{(n)}}{(-2\ell)_{(n)}} \frac{\left( \frac{r}{2M} \right)^{-n-k}}{n! k} \,.
\end{aligned}
\end{equation}
This term contains the correct power, with coefficient
\begin{equation}
\begin{aligned}
    &R(r) \supset \frac{M_{a_{2+} a_-}}{(r_+ - r_-)^{s+1}} \frac{r^{\ell-s}}{((2M)^{\ell+1}} i \frac{\omega-m\Omega}{4 \pi T_H} \left( \frac{r}{2M} \right)^{-2\ell-1} \bigg(  \sum_{n=0}^{\ell+s} \frac{(-\ell -s)_{(n)} (-\ell )_{(n)}}{(-2\ell)_{(n)} n! (2\ell+1-n)} + \\ &+s\sum_{n=0}^{\ell+s} \frac{(-\ell -s)_{(n)} (-\ell )_{(n)}}{(-2\ell)_{(n)} n! (2\ell-n)} + \frac{s(s+1)}{2} \sum_{n=0}^{\ell+s} \frac{(-\ell -s)_{(n)} (-\ell )_{(n)}}{(-2\ell)_{(n)} n! (2\ell-1-n)} \bigg) \,.
\end{aligned}
\end{equation}
A surprising identity reveals that
\begin{equation}
\begin{aligned}
    &\sum_{n=0}^{\ell+s} \frac{(-\ell -s)_{(n)} (-\ell )_{(n)}}{(-2\ell)_{(n)} n! (2\ell+1-n)} +s \sum_{n=0}^{\ell+s} \frac{(-\ell -s)_{(n)} (-\ell )_{(n)}}{(-2\ell)_{(n)} n! (2\ell-n)} + \frac{s(s+1)}{2} \sum_{n=0}^{\ell+s} \frac{(-\ell -s)_{(n)} (-\ell )_{(n)}}{(-2\ell)_{(n)} n! (2\ell-1-n)} = \\ &= \frac{\left( \ell+s \right)! \left( \ell-s \right)! \left( \ell ! \right)^2}{\left( 2 \ell \right)! \left( 2\ell+1 \right)!} \left( -1 \right)^s \,,
\end{aligned}
\end{equation}
therefore 
\begin{equation}
    R(r) \supset \frac{M_{a_{2+} a_-}}{(r_+ - r_-)^{s+1}} \frac{r^{\ell-s}}{(2M)^{\ell+1}} \left[ 1+i \frac{\omega-m\Omega}{4 \pi T_H} \left( \frac{r}{2M} \right)^{-2\ell-1} \frac{\left( \ell+s \right)! \left( \ell-s \right)! \left( \ell ! \right)^2}{\left( 2 \ell \right)! \left( 2\ell+1 \right)!} \left( -1 \right)^s \right] \,.
\end{equation}
Noticing that $1/4 \pi T_H \simeq 2 M$ finally gives the Love number 
\begin{equation}
    k_{a \,, m}^s = 2 i M \left( \omega - m \Omega \right) \left( -1 \right)^s \frac{\left( \ell+s \right)! \left( \ell-s \right)! \left( \ell ! \right)^2}{\left( 2 \ell \right)! \left( 2\ell+1 \right)!}   + \mathcal{O}(M^2\omega^2,M^2\Omega^2,M^2\omega \Omega)\,.
\label{eq:LNfinal}
\end{equation}
This result matches with formula (6.17) in \cite{charalambous2021vanishing}. Note that the Love number remains purely imaginary for a small frequency perturbation, and that it vanishes in the case of a static perturbation of a Schwarzschild black hole.

\vspace{1cm}
\textbf{Acknowledgements}: We would like to thank M. Bianchi, E. Franzin, A. Grassi, O. Lisovyy and J.F. Morales for fruitful discussions. This research is partially supported by the INFN Research Projects GAST and ST\&FI, by PRIN "Geometria delle varietà algebriche" and by PRIN "Non-perturbative Aspects Of Gauge Theories And Strings".

\appendix
\section{The radial and angular potentials}\label{coefficients_potential}
Both the radial and angular part of the Teukolsky equation can be written as a Schr\"odinger equation:
\begin{equation}
\frac{d^2 \psi(z)}{dz^2} + V(z)\psi(z) = 0
\end{equation}
with potential
\begin{equation}
V(z) = \frac{1}{z^2 (z-1)^2}\sum_{i=0}^4 \hat{A}_i z^i\,.
\end{equation}
For the radial part, the coefficients are given by
\begin{equation}\label{radial_As}
\begin{aligned}
&\hat{A}^r_0 = \frac{\text{a}^2(1-m^2) - M^2 + 4\text{a} m M \omega (M - \sqrt{ M^2 -\text{a}^2 }) + 4 M^2 \omega^2 (a^2 - 2 M^2) + 8 M^3 \sqrt{M^2-\text{a}^2} \omega^2}{4(\text{a}^2-M^2)} + \\&+ (i s) \frac{\text{a} m \sqrt{M^2-\text{a}^2} - 2 \text{a}^2 M \omega + 2 M^2 \omega (M - \sqrt{M^2 - \text{a}^2})}{2 (\text{a}^2 - M^2)} - \frac{s^2}{4} \,, \\
& \hat{A}^r_1 = \frac{4\text{a}^2 \lambda - 4 M^2 \lambda + (8\text{a} m M \omega+ 16\text{a}^2 M \omega^2 - 32 M^3 \omega^2) \sqrt{M^2-\text{a}^2}  + 4\text{a}^4 \omega^2 - 36\text{a}^2 M^2 \omega^2 + 32 M^4 \omega^2  }{4(\text{a}^2-M^2)} + \\ &+ (is) \left( - i + \frac{(2 \text{a}^2 \omega- \text{a} m )\sqrt{M^2 - \text{a}^2} }{\text{a}^2 - M^2} \right) + s^2 \,, \\
& \hat{A}^r_2 = -\lambda - 5\text{a}^2 \omega^2 + 12 M^2 \omega^2 - 12 M \omega^2 \sqrt{M^2-\text{a}^2} + (is) (i - 6 \omega \sqrt{M^2-\text{a}^2}) - s^2 \,, \\
& \hat{A}^r_3 = 8\text{a}^2 \omega^2 - 8 M^2 \omega^2 + 8 M \omega^2 \sqrt{M^2-\text{a}^2} + (is) 4 \omega \sqrt{M^2 - \text{a}^2} \,, \\
& \hat{A}^r_4 = 4 (M^2-\text{a}^2) \omega^2 \,,
\end{aligned}
\end{equation}
while for the angular part they are
\begin{equation}
\begin{aligned}
&\hat{A}^\theta_0 = - \frac{1}{4} (-1+m-s) (1+m-s) \,, \\
&\hat{A}^\theta_1 = c^2 + s + 2 c s - m s + s^2 + \lambda \,, \\
&\hat{A}^\theta_2 = - s - (c+s)(5c+s) - \lambda \,, \\
&\hat{A}^\theta_3 = 4c (2c+s) \,, \\
&\hat{A}^\theta_4 = - 4 c^2 \,.
\end{aligned}
\label{eq:angkerrA's}
\end{equation}

\section{CFT calculations}\label{CFT_calculations}
\subsection{The BPZ equation}\label{TheBPZequation}
To calculate the BPZ equation for the correlator (\ref{correlator_unnormalized}) we first evaluate the correlator with an extra insertion of the energy-momentum tensor:
\begin{equation}
\begin{aligned}
    & \langle \Delta, \Lambda_0, m_0 |T(w) \Phi_{2,1}(z) V_2(y) | V_1 \rangle = \\ 
    & = \sum_{n\geq0} \frac{1}{w^{n+2}} \langle \Delta, \Lambda_0, m_0 | [L_n,\Phi_{2,1}(z) V_2(y)] | V_1 \rangle + \bigg(\frac{\Delta_1}{w^2} + \frac{m_0 \Lambda_0}{w} + \Lambda_0^2\bigg) \langle \Delta, \Lambda_0, m_0 | \Phi_{2,1}(z) V_2(y) | V_1 \rangle = \\
    & = \bigg( \frac{z}{w} \frac{1}{w-z} \partial_z + \frac{\Delta_{2,1}}{(w-z)^2} + \frac{y}{w} \frac{1}{w-y} \partial_y + \frac{\Delta_2}{(w-y)^2} + \frac{\Delta_1}{w^2} + + \frac{m_0 \Lambda_0}{w} + \Lambda_0^2\bigg) \langle \Delta, \Lambda_0, m_0 | \Phi_{2,1}(z) V_2(y) | V_1 \rangle\,.
\end{aligned}
\end{equation}
Now we can simply compute
\begin{equation}
\begin{aligned}
    & \langle \Delta, \Lambda_0, m_0 | L_{-2} \cdot \Phi_{2,1}(z) V_2(y) | V_1 \rangle = \oint_{C_z} \frac{dw}{w-z} \langle \Delta, \Lambda_0, m_0 | T(w) \Phi_{2,1}(z) V_2(y) | V_1 \rangle = \\
    = &  \bigg( - \frac{1}{z}\partial_z + \frac{y}{z} \frac{1}{z-y} \partial_y + \frac{\Delta_2}{(z-y)^2} + \frac{\Delta_1}{z^2} + \frac{m_0 \Lambda_0}{z} + \Lambda_0^2\bigg) \langle \Delta, \Lambda_0, m_0 | \Phi_{2,1}(z) V_2(y) | V_1 \rangle\,.
\end{aligned}
\end{equation}
Using the Ward identity for $L_0$:
\begin{equation}
    \big( z\partial_z + y\partial_y - \Lambda_0 \partial_{\Lambda_0} + \Delta_{2,1} + \Delta_2 + \Delta_1 - \Delta \big) \langle \Delta, \Lambda_0, m_0 | \Phi_{2,1}(z) V_2(y) | V_1 \rangle = 0
\end{equation}
we can eliminate $\partial_y$. Then setting $y=1$ we obtain
\begin{equation}
\begin{aligned}\label{L-2action}
    & \langle \Delta, \Lambda_0, m_0 | L_{-2} \cdot \Phi_{2,1}(z) V_2(1) | V_1 \rangle = \\
    = & \bigg( - \frac{1}{z}\partial_z - \frac{1}{z} \frac{1}{z-1} \big(z\partial_z - \Lambda_0 \partial_{\Lambda_0} + \Delta_{2,1} + \Delta_2 + \Delta_1 - \Delta \big) + \frac{\Delta_2}{(z-1)^2} + \frac{\Delta_1}{z^2} + \frac{m_0 \Lambda_0}{z} + \Lambda_0^2\bigg) \Psi(z) 
\end{aligned}
\end{equation}
which gives the BPZ equation
\begin{equation}
\begin{aligned}
    0 =& \langle \Delta, \Lambda_0, m_0 | \big(b^{-2} \partial_z^2 + L_{-2}\cdot \big) \Phi_{2,1}(z) V_2(1) | V_1 \rangle = \\
    = & \bigg(b^{-2} \partial_z^2 - \frac{1}{z}\partial_z - \frac{1}{z} \frac{1}{z-1} \big(z\partial_z - \Lambda_0 \partial_{\Lambda_0} + \Delta_{2,1} + \Delta_2 + \Delta_1 - \Delta \big) + \frac{\Delta_2}{(z-1)^2} + \frac{\Delta_1}{z^2} + \frac{ m_0 \Lambda_0}{z} + \Lambda_0^2\bigg) \Psi(z) \,.
\end{aligned}
\end{equation}
\subsection{DOZZ factors}\label{DOZZfactors}
We normalize vertex operators so that the DOZZ three-point function\cite{Dorn:1994xn, Zamolodchikov:1995aa} reads
\begin{equation}
C\left(\alpha_1,\alpha_2,\alpha_3\right) =  \frac{1}{\Upsilon_b(\alpha_1+\alpha_2+\alpha_3 + \frac{Q}{2}) \Upsilon_b(\alpha_1+\alpha_2-\alpha_3 + \frac{Q}{2})\Upsilon_b(\alpha_2+\alpha_3-\alpha_1 + \frac{Q}{2})\Upsilon_b(\alpha_3+\alpha_1-\alpha_2 + \frac{Q}{2})} \,,
\label{eq:DOZZ}
\end{equation}
where 
\begin{equation}
\begin{aligned}
&\Upsilon_b (x) = \frac{1}{\Gamma_b(x) \Gamma_b(Q-x)} \,, \\
&\Gamma_b(x) = \frac{\Gamma_2(x | b, b^{-1})}{\Gamma_2(\frac{Q}{2} | b, b^{-1})} \,,
\end{aligned}
\end{equation}
and $\Gamma_2$ is the double gamma function. $\U$ satisfies the shift relation
\begin{equation}
\Upsilon_b (x + b) = \gamma (bx) b^{1- 2 b x} \Upsilon_b (x) \,.
\label{eq:shiftU}
\end{equation}
Moreover $\gamma(x) = \G (x) / \G (1-x)$, and satisfies the following relations
\begin{equation}
\begin{aligned}
&\gamma (- x) \gamma (x) = -\frac{1}{x^2} \,, \\
&\gamma (x + 1) = -x^2 \gamma (x) \,, \\
&\gamma(x) = \frac{1}{\gamma(1-x)} \,.
\end{aligned}
\label{eq:sgammaprop}
\end{equation}
The two-point function normalization is given in terms of the DOZZ factors, that is
\begin{equation}
\langle \Delta_\alpha | \Delta_\alpha \rangle = G(\alpha) = C(\alpha, - \frac{Q}{2}, \alpha) = \frac{1}{\Upsilon_b(0) \Upsilon_b(0)\Upsilon_b(2 \alpha)\Upsilon_b(2 \alpha + Q)} \,.
\end{equation}
The regular OPE coefficient appearing in section \ref{section:ConnectionProblem} can be explicitly computed in terms of DOZZ factors, that is
\begin{equation}
\mathcal{C}_{\alpha_{2, 1}, \alpha_i}^{\alpha_{i \pm}} = G^{-1}(\alpha_{i \pm}) C(\alpha_{i \pm}, \frac{-b-Q}{2}, \alpha_i) = \gamma (- b^2) \gamma(\mp 2 b \alpha_i) b^{2 b (\pm 2 \alpha + Q)} \,.
\end{equation}
Another relevant ratio is 
\begin{equation}
\frac{C (\alpha_1, \alpha_2, \alpha_{3+})}{C (\alpha_1, \alpha_2, \alpha_{3-})} = b^{- 8 b \alpha_3} \prod_{\pm, \pm} \gamma(\frac{1}{2} + b (\pm \alpha_1 \pm \alpha_2 + \alpha_3)) \,,
\end{equation}
that is readily computed from the shift relation (\ref{eq:shiftU}). With these relations at our disposal, we can evaluate ratios of the $K$s appearing in equations (\ref{eq:connectionconstraints}). In particular,
\begin{equation}
\frac{K^{(t)}_{\alpha_{2-}, \alpha_{2-}}}{K^{(t)}_{\alpha_{2+}, \alpha_{2+}}} = \frac{G^{-1}(\alpha_{2-}) C(\alpha_{2-}, \frac{-b-Q}{2}, \alpha_2) C(\alpha, \alpha_{2-}, \alpha_1)}{G^{-1}(\alpha_{2+}) C(\alpha_{2+}, \frac{-b-Q}{2}, \alpha_2) C(\alpha, \alpha_{2+}, \alpha_1)} = \frac{\gamma(2 b \alpha_2)}{\gamma(-2b\alpha_2)} \prod_{\pm, \pm} \gamma(\frac{1}{2} + b (\pm \alpha \pm \alpha_1 - \alpha_2)) \,,
\end{equation}
and similarly
\begin{equation}
\frac{K^{(u)}_{\alpha_+, \alpha_+}}{K^{(u)}_{\alpha_-, \alpha_-}} = \frac{\gamma(-2 b \alpha)}{\gamma(2b\alpha)} \prod_{\pm, \pm} \gamma(\frac{1}{2} + b ( \alpha \pm \alpha_1 \pm \alpha_2)) \,.
\end{equation}
\subsection{Irregular OPE}\label{IrregularOPE}
Following \cite{Gaiotto:2012sf} let us make the following Ansatz for the OPE with the irregular state 
\begin{equation}\label{irrAnsatz}
\langle \Delta_\alpha, \Lambda_0, \bar{\Lambda}_0, m_0 | \Phi_{2, 1} (z, \bar{z}) = \sum_{\beta} \tilde{\mathcal{C}}^{\beta}_{\alpha, \alpha_{2,1}} \displaystyle\left\lvert \sum_{\mu_0, k} \mathcal{A}_{\beta, \mu_0} z^{\zeta} \Lambda_0^{\lambda} e^{\gamma \Lambda_0 z} z^{-k} \langle \Delta_\beta, \Lambda_0, \mu_0; k | \right\rvert^2 \,,
\end{equation}
with all the parameters to be determined. Here $\langle \Delta_\beta, \Lambda_0, \mu_0, k |$ is the $k$-th irregular descendant, that schematically has the form
\begin{equation}
| \Delta_\beta, \Lambda_0, \mu_0; k \rangle \sim \sum L_{-J} \Lambda_0^{-k''} \partial_{\Lambda_0}^{k'} | \Delta_\beta, \Lambda_0, \mu_0 \rangle 
\end{equation}
where the sum runs over all $k',\,k'',\, J$ such that $k' + k'' + | J | = k$, with appropriate coefficients that can be determined from the Ward identities. Note that in principle the parameters $\zeta, \lambda, \gamma$ depend both on $\beta$ and $\mu_0$. The first constraint comes from comparing with the regular OPE, namely
\begin{equation}
\langle \Delta_\alpha, \Lambda_0, \bar{\Lambda}_0, m_0 | \Phi_{2, 1} (z, \bar{z}) | \Delta_\beta \rangle \sim \langle \Delta_\alpha, \Lambda_0, \bar{\Lambda}_0, \mu_0 | \Delta_{\beta_\pm} \rangle =  \langle \Delta_\alpha | \Delta_{\beta_\pm} \rangle \Rightarrow \beta_\pm = \alpha\,.
\label{eq:IRRfirstconstraint}
\end{equation}
The other coefficients can be fixed by acting with the Virasoro generators on the left and right hand sides of the Ansatz (\ref{irrAnsatz}). Focusing on the chiral correlator and comparing powers of $\Lambda$ and $z$ we have 
\begin{equation}
\begin{aligned}
\langle \Delta_\alpha, \Lambda_0, m_0 | \Phi_{2, 1} (z) L_0 &= (\Delta_{\alpha} + \Lambda_0 \partial_{\Lambda_0} - \Delta_{2, 1} - z \partial_z) \langle \Delta_\alpha, \Lambda_0, m_0 | \Phi_{2, 1} (z) = \\ &= \sum_{k} z^{\zeta - k} \Lambda_0^\lambda e^{\gamma \Lambda_0 z} (\Delta_\alpha - \Delta_{2,1} - \zeta + k + \lambda + \Lambda_0 \partial_{\Lambda_0}) \langle \Delta_\beta, \Lambda_0, \mu_0; k | = \\ &= \sum_{k} z^{\zeta - k} \Lambda_0^\lambda e^{\gamma \Lambda_0 z} (\Delta_\beta + k + \Lambda_0 \partial_{\Lambda_0}) \langle \Delta_\beta, \Lambda_0, \mu_0; k | \,,
\end{aligned}
\end{equation}
that gives the constraint
\begin{equation}
\lambda - \zeta = \Delta_\beta - \Delta_\alpha + \Delta_{2, 1} \,.
\end{equation}
Now let us consider the action of $L_{-1}$. We have
\begin{equation}
\begin{aligned}
\langle \Delta_\alpha, \Lambda_0, m_0 | \Phi_{2, 1} (z) L_{-1} &= (m_0 \Lambda_0 - \partial_z) \langle \Delta_\alpha, \Lambda_0, m_0 | \Phi_{2, 1} (z) = \\ &= \sum_k z^{\zeta} \Lambda_0^\lambda e^{\gamma \Lambda_0 z} ((m_0  - \gamma )\Lambda_0z^{-k} + (k - \zeta)z^{-k-1} ) \langle \Delta_\beta, \Lambda_0, \mu_0; k | = \\ &= z^{\zeta} \Lambda_0^\lambda e^{\gamma \Lambda_0 z} \left( \langle \Delta_\beta, \Lambda_0, \mu_0| \mu_0 \Lambda_0 + z^{-1} \langle \Delta_\beta, \Lambda_0, \mu_0; 1| L_{-1} + \dots \right)\,.
\end{aligned}
\end{equation}
Comparing powers,
\begin{equation}
\begin{aligned}
&\mathcal{O} (z^\zeta) \Rightarrow  m_0 - \gamma = \mu_0 \,, \\
&\mathcal{O} (z^{\zeta-1}) \Rightarrow \mu_0 \Lambda_0 \langle \Delta_\beta, \Lambda_0, \mu_0; 1| - \zeta \langle \Delta_\beta, \Lambda_0, \mu_0| = \langle \Delta_\beta, \Lambda_0, \mu_0; 1| L_{-1} \,.
\end{aligned}
\label{eq:IRRsecondconstraint}
\end{equation}
The first irregular descendant is of the form\footnote{The term $\sim \Lambda^{-1}$ cannot be determined at this order. Luckily, it doesn't play any role in the following discussion.}
\begin{equation}
\langle \Delta_\beta, \Lambda_0, \mu_0; 1| = A \langle \Delta_\beta, \Lambda_0, \mu_0 | L_{1} + B \partial_{\Lambda_0} \langle \Delta_\beta, \Lambda_0, \mu_0 | \,,
\end{equation}
therefore equation (\ref{eq:IRRsecondconstraint}) gives 
\begin{equation}
\begin{aligned}
\mu_0 \Lambda_0 \left( A \langle \Delta_\beta, \Lambda_0, \mu_0 | L_{1} + B \partial_{\Lambda_0} \langle \Delta_\beta, \Lambda_0, \mu_0 | \right) - \zeta \langle \Delta_\beta, \Lambda_0, \mu_0 | = A \langle \Delta_\beta, \Lambda_0, \mu_0 | L_{1} L_{-1}+ B \partial_{\Lambda_0} \langle \Delta_\beta, \Lambda_0, \mu_0 | L_{-1} \,.
\end{aligned}
\end{equation}
The RHS gives
\begin{equation}
\begin{aligned}
&A (2 \Delta_\beta + 2 \Lambda_0 \partial_{\Lambda_0}) \langle \Delta_\beta, \Lambda_0, \mu_0| +  A \mu_0 \Lambda_0 \langle \Delta_\beta, \Lambda_0, \mu_0| L_1 + B \mu_0 \langle \Delta_\beta, \Lambda_0, \mu_0| + B \mu_0 \Lambda_0 \partial_{\Lambda_0} \langle \Delta_\beta, \Lambda_0, \mu_0| = \\ &= (2 A \Delta_\beta + B \mu_0) \langle \Delta_\beta, \Lambda_0, \mu_0| + (2 A \Lambda_0 + B \mu_0 \Lambda_0) \partial_{\Lambda_0} \langle \Delta_\beta, \Lambda_0, \mu_0| + A \mu_0 \Lambda_0 \langle \Delta_\beta, \Lambda_0, \mu_0| L_1 \,.
\end{aligned}
\end{equation}
Comparing term by term, we obtain equations for $A, B$
\begin{equation}
\begin{aligned}
&2 A \Delta_\beta + B \mu_0 = - \zeta \,, \\
&2 A \Lambda_0 + B \mu_0 \Lambda_0 = B \mu_0 \Lambda_0 \,, \\
&\Rightarrow A = 0 \,, \, B = - \frac{\zeta}{\mu_0} \,.
\end{aligned}
\end{equation}
Another constraint comes from the action of $L_2$. We have
\begin{equation}
\begin{aligned}
\langle \Delta_\alpha, \Lambda_0, m_0 | \Phi_{2, 1} (z) L_{-2} &= (\Lambda_0^2 - z^{-1} \partial_z + \Delta_{2, 1} z^{-2}) \langle \Delta_\alpha, \Lambda_0, m_0 | \Phi_{2, 1} (z) = \\ &= \sum_k z^{\zeta} \Lambda_0^\lambda e^{\gamma \Lambda_0 z} (\Lambda_0^2 z^{-k}  - \gamma \Lambda_0z^{-k-1} + (k - \zeta + \Delta_{2,1})z^{-k-2} ) \langle \Delta_\beta, \Lambda_0, \mu_0; k | = \\ &= z^{\zeta} \Lambda_0^\lambda e^{\gamma \Lambda_0 z} \left( \langle \Delta_\beta, \Lambda_0, \mu_0| \Lambda_0^2 + z^{-1} \langle \Delta_\beta, \Lambda_0, \mu_0; 1| L_{-2} + \dots \right) = \\ &= z^{\zeta} \Lambda_0^\lambda e^{\gamma \Lambda_0 z} \left( \langle \Delta_\beta, \Lambda_0, \mu_0| \Lambda_0^2 - z^{-1} \frac{\zeta}{\mu_0} (2 \Lambda_0 + \Lambda_0^2 \partial_{\Lambda_0}) \langle \Delta_\beta, \Lambda_0, \mu_0| + \dots \right)
\end{aligned}
\end{equation}
The previous equation is trivially satisfied at order $\mathcal{O}(z^\zeta)$, and comparing at order $\mathcal{O}(z^{\zeta-1})$ gives
\begin{equation}
(- \Lambda_0^2 \frac{\zeta}{\mu_0} \partial_{\Lambda_0} - \gamma \Lambda_0 ) \langle \Delta_\beta, \Lambda_0, \mu_0| = -\frac{\zeta}{\mu_0} (2 \Lambda_0 + \Lambda_0^2 \partial_{\Lambda_0}) \langle \Delta_\beta, \Lambda_0, \mu_0| \,,
\end{equation}
that finally gives
\begin{equation}
\gamma = 2 \frac{\zeta}{\mu_0} \,.
\end{equation}
The last constraint we need is most easily obtained by looking at the null-state equation satisfied by the irregular 3 point function (\ref{eq:IRRfirstconstraint}). We have 
\begin{equation}
\langle \Delta_\alpha, \Lambda_0, m_0 | T(w) \Phi_{2, 1} (z) | \Delta_{\alpha_\pm} \rangle = \langle \Delta_\alpha, \Lambda_0, m_0 | \left( \frac{m_0 \Lambda_0}{w} + \Lambda_0^2 + \frac{\Delta_{\alpha_\pm}}{w^2} +  \frac{\Delta_{2,1}}{(w-z)^2} + \frac{z/w}{w-z} \partial_z \right) \Phi_{2, 1} (z) | \Delta_{\alpha_\pm} \rangle \,,
\end{equation}
therefore
\begin{equation}
\left( b^{-2} \partial_z^2 - \frac{1}{z} \partial_z + \frac{\Delta_{\alpha \pm}}{z^2} + \frac{m_0 \Lambda_0}{z} + \Lambda_0^2 \right) \langle \Delta_\alpha, \Lambda_0, m_0 | \Phi_{2, 1} (z) | \Delta_{\alpha_\pm} \rangle = 0 \,.
\end{equation}
Substituting the irregular OPE and looking at the leading term as $z \to \infty$ gives
\begin{equation}
\left( \frac{\gamma \Lambda_0}{b} \right)^2 + \Lambda_0^2 = 0 \Rightarrow \gamma = \pm i b \,.
\end{equation}
Putting all the constraints together yields, for a fixed channel $\beta = \alpha_\theta$, $\theta = \pm$,
\begin{equation}
\begin{aligned}
&\gamma = \pm i b \,, \\
&\zeta = \frac{1}{2} \left( b^2 \pm i b m_0 \right) = \frac{1}{2} \left( bQ - 1 \pm 2 m_3 \right) \,, \\
&\lambda - \zeta = - \frac{1}{2} bQ + \theta b \alpha_\theta \,, \\ 
&\mu_0 = m_{0 \pm} = m_0 \pm (- i b) \,.
\end{aligned}
\end{equation}
Finally, the irregular OPE reads
\begin{equation}
\begin{aligned}
\langle \Delta_\alpha, \Lambda_0, \bar{\Lambda}_0, m_0 | \Phi_{2, 1} (z, \bar{z}) &=  \tilde{\mathcal{C}}^{\alpha_+}_{\alpha, \alpha_{2,1}} \displaystyle\left\lvert \sum_{\pm, k} \mathcal{A}_{\alpha_+, m_{0 \pm}} \Lambda^{-\frac{1}{2} b Q + b \alpha_+} (\Lambda z)^{\frac{1}{2} (bQ - 1 \pm 2 m_3)} e^{\pm \Lambda z/2} z^{-k} \langle \Delta_{\alpha_+}, \Lambda_0, m_{0 \pm}; k | \right\rvert^2 + \\ &+ \tilde{\mathcal{C}}^{\alpha_-}_{\alpha, \alpha_{2,1}} \displaystyle\left\lvert \sum_{\pm, k} \mathcal{A}_{\alpha_-, m_{0 \pm}} \Lambda^{-\frac{1}{2} b Q - b \alpha_-} (\Lambda z)^{\frac{1}{2} (bQ - 1 \pm 2 m_3)} e^{\pm \Lambda z/2} z^{-k} \langle \Delta_{\alpha_-}, \Lambda_0, m_{0 \pm}; k | \right\rvert^2\,.
\end{aligned}
\end{equation}
where we absorbed a $2 i b$ factor in the OPE coefficients for later convenience, $\Lambda = 2ib \Lambda_0$ and $m_3 = \frac{i}{2}b m_0 $. Here the irregular state depending on $\Lambda_0, \bar{\Lambda}_0$ denotes the full (chiral$\otimes$antichiral) state, and the modulus squared of the chiral states (depending only on $\Lambda_0$) has to be understood as a tensor product. Now we can fix the OPE coefficients $\tilde{\mathcal{C}}^{\alpha_\pm}_{\alpha, \alpha_{2,1}}, \mathcal{A}_{\alpha_\pm, m_{0 \pm}}$ making use of the NSE for the full irregular three point function. Namely, 
\begin{equation}
\left( b^{-2} \partial_z^2 - \frac{1}{z} \partial_z + \frac{\Delta_{\alpha \pm}}{z^2} + \frac{m_0 \Lambda_0}{z} + \Lambda_0^2 \right) \langle \Delta_\alpha, \Lambda_0, \bar{\Lambda}_0, m_0 | \Phi_{2, 1} (z, \bar{z}) | \Delta_{\alpha_\pm} \rangle = 0 \,.
\end{equation}
If we define $\langle \Delta_\alpha, \Lambda_0, \bar{\Lambda}_0, m_0 | \Phi_{2, 1} (z, \bar{z}) | \Delta_{\alpha_\pm} \rangle = \displaystyle\left\lvert e^{- \frac{\Lambda z}{2}} (\Lambda z)^{\frac{1}{2} (bQ + 2 b \alpha_\pm)} \right\rvert^2 G_\pm (z, \bar{z})$, then $G(z, \bar{z})$ will satisfy
\begin{equation}
\left( z \partial_z^2 + (1 + 2 b \alpha_\pm - \Lambda z ) \partial_z - \frac{\Lambda}{2} (1 + 2 m_3 + 2 b \alpha_\pm) \right) G_\pm (z, \bar{z}) = 0 \,.
\end{equation}
Note that we can rewrite the previous equation using the natural variable $w = \Lambda z$, and obtain
\begin{equation}
\left( w \partial_w^2 + (1 + 2 b \alpha_\pm - w ) \partial_w - \frac{1}{2} (1 + 2 m_3 + 2 b \alpha_\pm) \right) G_\pm (w, \bar{w}) = 0 \,.
\label{eq:GNSE}
\end{equation}
Equation (\ref{eq:GNSE}) is the confluent hypergeometric equation, therefore\footnote{Note that in principle also mixed terms could appear. However, they cannot be there in order to correctly match the behavior near zero.}
\begin{equation}
G_\pm (w, \bar{w}) = K_\pm^{(1)}  \displaystyle\left\lvert {}_1 F_1 \left( \frac{1}{2} + m_3 + b \alpha_\pm, 1 + 2 b \alpha_\pm, w \right) \right\rvert^2 + K_\pm^{(2)}  \displaystyle\left\lvert w^{-2 b \alpha_\pm} {}_1 F_1 \left( \frac{1}{2} + m_3 - b \alpha_\pm, 1 - 2 b \alpha_\pm, w \right) \right\rvert^2 \,.
\label{eq:GNSEsol}
\end{equation}
Expanding the correlator near zero and comparing the solution (\ref{eq:GNSEsol}) with the regular OPE near zero,
\begin{equation}
K_\pm^{(1)} \displaystyle\left\lvert (\Lambda z )^{\frac{1}{2} b Q + b \alpha_\pm} \right\rvert^2 + K_\pm^{(2)} \displaystyle\left\lvert (\Lambda z )^{\frac{1}{2} b Q - b \alpha_\pm} \right\rvert^2 = G(\alpha) \mathcal{C}^{\alpha}_{\alpha_{2,1} \alpha_{\pm}} \displaystyle\left\lvert z^{\frac{1}{2} b Q \mp b \alpha_\pm} \right\rvert^2 \,,
\end{equation}
and hence
\begin{equation}
\begin{aligned}
&K_+^{(1)} = 0 \,, \, K_+^{(2)} = G(\alpha) \mathcal{C}^{\alpha}_{\alpha_{2,1} \alpha_+} \displaystyle\left\lvert \Lambda^{-\frac{1}{2} b Q + b \alpha_+} \right\rvert^2 \,, \\
&K_-^{(1)} = G(\alpha) \mathcal{C}^{\alpha}_{\alpha_{2,1} \alpha_-} \displaystyle\left\lvert \Lambda^{-\frac{1}{2} b Q - b \alpha_-} \right\rvert^2 \,, \, K_-^{(2)} = 0\,.
\end{aligned}
\end{equation}
Now expanding the confluent hypergeometric near infinity and matching with the OPE we can finally fix all the coefficients. Recall that as $w \to \infty$
\begin{equation}
\begin{aligned}
{}_1 F_1 \left( \frac{1}{2} + m_3 + b \alpha_\pm, 1 + 2 b \alpha_\pm, w \right) &\simeq \frac{\Gamma (1 + 2 b \alpha_\pm)}{\Gamma (\frac{1}{2} + m_3 + b \alpha_\pm)} e^w w^{-\frac{1}{2} + m_3 - b \alpha_\pm} + \\ &+ \frac{\Gamma (1 + 2 b \alpha_\pm)}{\Gamma (\frac{1}{2} - m_3 + b \alpha_\pm)} (- 1)^{-\frac{1}{2} - m_3 - b \alpha_\pm} (w )^{-\frac{1}{2} - m_3 - b \alpha_\pm} \,, \\
w^{-2 b \alpha_\pm} {}_1 F_1 \left( \frac{1}{2} + m_3 - b \alpha_\pm, 1 - 2 b \alpha_\pm, w \right) &\simeq \frac{\Gamma (1 - 2 b \alpha_\pm)}{\Gamma (\frac{1}{2} + m_3 - b \alpha_\pm)} e^w w^{-\frac{1}{2} + m_3 - b \alpha_\pm} + \\ &+ \frac{\Gamma (1 - 2 b \alpha_\pm)}{\Gamma (\frac{1}{2} - m_3 - b \alpha_\pm)} (- 1)^{-\frac{1}{2} - m_3 + b \alpha_\pm} (w )^{-\frac{1}{2} - m_3 - b \alpha_\pm} \,.
\end{aligned}
\end{equation}
Let us concentrate on the $\alpha_+$ channel. Expanding the full correlator and matching $z$ powers gives
\begin{equation}
\begin{aligned}
&G(\alpha) \mathcal{C}^{\alpha}_{\alpha_{2,1} \alpha_+} \displaystyle\left\lvert \Lambda^{-\frac{1}{2} b Q + b \alpha_+} \right\rvert^2 \times \\ & \times \displaystyle\left\lvert \frac{\Gamma (1 - 2 b \alpha_+)}{\Gamma (\frac{1}{2} + m_3 - b \alpha_+)} e^{w/2} w^{\frac{bQ}{2}-\frac{1}{2} + m_3} + \frac{\Gamma (1 - 2 b \alpha_+)}{\Gamma (\frac{1}{2} - m_3 - b \alpha_+)} (- 1)^{-\frac{1}{2} - m_3 + b \alpha_+} e^{-w/2} (w )^{\frac{bQ}{2}-\frac{1}{2} - m_3} \right\rvert^2 = \\ &=  G(\alpha_+) \tilde{\mathcal{C}}^{\alpha_+}_{\alpha, \alpha_{2,1}} \displaystyle\left\lvert \sum_{\pm} \mathcal{A}_{\alpha_+, m_{0 \pm}} \Lambda^{-\frac{1}{2} b Q + b \alpha_+} (\Lambda z)^{\frac{1}{2} (bQ - 1 \pm 2 m_3)} e^{\pm \Lambda z/2} \right\rvert^2 \,.
\end{aligned}
\label{eq:IRROPEMatching}
\end{equation}
Finally from equation (\ref{eq:IRROPEMatching}) we can read off the coefficients (the coefficients for the $\alpha_-$ channel are obtained simply sending $\alpha_+ \to - \alpha_-$)
\begin{equation}
\begin{aligned}
&\tilde{\mathcal{C}}^{\alpha_\pm}_{\alpha, \alpha_{2,1}} = \mathcal{C}^{\alpha_\pm}_{\alpha_{2,1} \alpha} \,, \\
&\mathcal{A}_{\alpha_+, m_{0 +}} = \frac{\Gamma (1 - 2 b \alpha_+)}{\Gamma (\frac{1}{2} + m_3 - b \alpha_+)} \,, \\
&\mathcal{A}_{\alpha_+, m_{0 -}} =  \frac{\Gamma (1 - 2 b \alpha_+)}{\Gamma (\frac{1}{2} - m_3 - b \alpha_+)} (- 1)^{-\frac{1}{2} - m_3 + b \alpha_+} \,, \\
&\mathcal{A}_{\alpha_-, m_{0 +}} = \frac{\Gamma (1 + 2 b \alpha_-)}{\Gamma (\frac{1}{2} + m_3 + b \alpha_-)} \,, \\
&\mathcal{A}_{\alpha_-, m_{0 -}} =  \frac{\Gamma (1 + 2 b \alpha_-)}{\Gamma (\frac{1}{2} - m_3 + b \alpha_-)} (- 1)^{-\frac{1}{2} - m_3 - b \alpha_-} \,.
\end{aligned}
\label{eq:irrOPEcoeff}
\end{equation}
Two remarks about equations (\ref{eq:irrOPEcoeff}): first of all, the OPE is symmetric in $\alpha \to - \alpha$, as it should be. Moreover, we expect the full irregular 3 point correlator to be symmetric under the simultaneous transformation $\Lambda \to - \Lambda, m_3 \to - m_3$. Under this transformation
\begin{equation}
\begin{aligned}
&\mathcal{A}_{\alpha_+, m_{3+}} \Lambda^{-\frac{1}{2} (bQ - 2 b \alpha_+)} e^{\frac{\Lambda z}{2}} \left( \Lambda z \right)^{\frac{1}{2} \left( bQ - 1 + 2 m_3 \right)} \to \\ & \to (-1)^{-\frac{1}{2} - m_3 +  b\alpha_+} \frac{\Gamma (1 - 2 b \alpha_+)}{\Gamma (\frac{1}{2} - m_3 - b \alpha_+)} \Lambda^{-\frac{1}{2} (bQ - 2 b \alpha_+)} e^{-\frac{\Lambda z}{2}} \left( \Lambda z \right)^{\frac{1}{2} \left( bQ - 1 - 2 m_3 \right)}  = \\ &= \mathcal{A}_{\alpha_+, m_{3-}} \Lambda^{-\frac{1}{2} (bQ - 2 b \alpha_+)} e^{-\frac{\Lambda z}{2}} \left( \Lambda z \right)^{\frac{1}{2} \left( bQ - 1 - 2 m_3 \right)} \,,
\end{aligned}
\end{equation}
and the same happens for the other channel. This suggests that the $(-1)^{-\frac{1}{2} - m_3 \pm b \alpha_ \pm}$ factor naturally multiplies $\Lambda$ in the irregular OPE. Therefore, after this minor change we obtain formulae (\ref{eq:ierrgope}), (\ref{eq:irrOPEcoeffreal}).

\section{Nekrasov formulae}\label{AppendixNekrasov}
\subsection{The AGT dictionary}
The irregular conformal blocks of the form $\langle \Delta_\alpha, \Lambda_0, m_0 | V_{\alpha_{2}} (1) | \Delta_{\alpha_1} \rangle$ can be efficiently computed as a gauge theory instanton partition function thanks to the AGT correspondence \cite{Alday_2010}. Concretely, we have
\begin{equation}
    \langle \Delta_\alpha, \Lambda_0, m_0 | V_{\alpha_{2}} (1) | \Delta_{\alpha_1} \rangle = \mathcal{Z}^{\mathrm{inst}}_{SU(2)}(\Lambda,a,m_1,m_2,m_3)
\end{equation}
where $\mathcal{Z}^{\mathrm{inst}}$ is the Nekrasov instanton partition function of $\mathcal{N}=2$ $SU(2)$ gauge theory with three hypermultiplets. The Nekrasov partition function contains a fundamental mass scale $\hbar = \sqrt{\epsilon_1\epsilon_2}$ which sets the units in which everything is measured. The mapping of parameters between CFT and gauge theory is then
\begin{equation}
    \begin{aligned}
    &\epsilon_1 = \frac{\hbar}{b}\, , \quad \epsilon_2 = \hbar b\, , \quad \epsilon=\epsilon_1+\epsilon_2 \quad  \longrightarrow\quad Q=\frac{\epsilon}{\hbar}\,, \\
    &\Delta_i = \frac{Q^2}{4}-\alpha_i^2 = \frac{\frac{\epsilon^2}{4}-a_i^2}{\hbar^2} \, , \quad a_i := \hbar\alpha_i\,, \\
    &\Lambda = 2i\hbar \Lambda_0 \, , \quad m_3 = \frac{i}{2}\hbar m_0\,,\\
    &m_1 = a_1 + a_2 \, , \quad m_2 = -a_1 + a_2\,.
    \end{aligned}
\end{equation}
The factors of $i\hbar$ in $\Lambda$ and $m_3$ do not appear in \cite{Marshakov_2009} where the irregular state is defined because they drop terms of the form $\sqrt{-\epsilon_1\epsilon_2}$. 
\subsection{The instanton partition function}
The $SU(2)$ partition function is given by the $U(2)$ partition function divided by the $U(1)$-factor:
\begin{equation}
    \mathcal{Z}^{\mathrm{inst}}_{SU(2)}(\Lambda,a,m_1,m_2,m_3,\epsilon_1,\epsilon_2)= \mathcal{Z}_{U(1)}^{-1}(\Lambda,m_1,m_2,\epsilon_1,\epsilon_2)\mathcal{Z}_{U(2)}^{\mathrm{inst}}(\Lambda,a,m_1,m_2,m_3,\epsilon_1,\epsilon_2)
\end{equation}
where the $U(2)$ partition function is given by a combinatorial formula which we review now. We often suppress the dependence on $\epsilon_1,\epsilon_2$. We mostly follow the notation of \cite{Alday_2010}. 

Let $Y=(\lambda_1 \geq \lambda_2 \geq ...)$ be a Young tableau where $\lambda_i$ is the height of the $i$-th column and we set $\lambda_i=0$ when $i$ is larger than the width of the tableau. Its transpose is denoted by $Y^T=(\lambda_1' \geq \lambda_2' \geq ...)$. For a box $s$ at the coordinate $(i,j)$ we define the arm-length $A_Y(s)$ and the leg-length $L_Y(s)$ with respect to the tableau $Y$ as 
\begin{equation}
    A_Y(s)=\lambda_i-j\,,\quad L_Y(s) = \lambda_j'-i\,.
\end{equation}
Note that they can be negative when $s$ is outside the tableau. Define a function $E$ by
\begin{equation}
    E(a,Y_1,Y_2,s) = a-\epsilon_1 L_{Y_2}(s)+\epsilon_2(A_{Y_1}(s)+1)\,.
\end{equation}
Using the notation $\Vec{a}=(a_1,a_2)$ with $a_1=-a_2=a$ and $\Vec{Y}=(Y_1,Y_2)$ the contribution of a vector multiplet is
\begin{equation}
    z^{\mathrm{inst}}_{\mathrm{vector}}(\Vec{a},\Vec{Y})=\prod_{i,j=1}^2 \prod_{s \in Y_i}\frac{1}{E(a_i-a_j,Y_i,Y_j,s)}\prod_{t \in Y_j}\frac{1}{\epsilon_1+\epsilon_2-E(a_j-a_i,Y_j,Y_i,t)}
\end{equation}
and that of an (antifundamental) hypermultiplet 
\begin{equation}
    z^{\mathrm{inst}}_{\mathrm{matter}}(\Vec{a},\Vec{Y},m)=\prod_{i=1}^2 \prod_{s \in Y_i} \left(a+m+\epsilon_1\left(i-\frac{1}{2}\right)+\epsilon_2\left(j-\frac{1}{2}\right)\right)\,.
\end{equation}
This is different from the formula given in \cite{Alday_2010} because our masses are shifted with respect to theirs by $\epsilon/2$. Finally, the $U(2)$ partition function is given by
\begin{equation}
    \mathcal{Z}_{U(2)}^{\mathrm{inst}}(\Lambda,a,m_1,m_2,m_3) = \sum_{\Vec{Y}}\Lambda^{|\Vec{Y}|} z^{\mathrm{inst}}_{\mathrm{vector}}(\Vec{a},\Vec{Y}) \prod_{n=1}^3 z^{\mathrm{inst}}_{\mathrm{matter}}(\Vec{a},\Vec{Y},m_n)\,,
\end{equation}
where $|\Vec{Y}|$ denotes the total number of boxes in $Y_1$ and $Y_2$. 

The $U(1)$-factor on the other hand can be obtained by decoupling one mass from the $U(1)$-factor for $N_f=4$. Before decoupling, the third and fourth masses are given by
\begin{equation}
    m_3 = a_3+a_4 \, , \quad m_4 = a_3 - a_4
\end{equation}
where $a_3$ and $a_4$ are related to the momenta of the two vertex operators that collide to form the irregular state. The $U(1)$-factor is
\begin{equation}
    \mathcal{Z}_{U(1)}^{N_f=4}=(1-q)^{2(a_2+\epsilon/2)(a_3+\epsilon/2)/\epsilon_1\epsilon_2} \,.
\end{equation}
The decoupling limit is given by $q \rightarrow 0 ,\, m_4 \rightarrow \infty$ with $q m_4 \equiv \Lambda$ finite. This gives the $N_f=3$ $U(1)$-factor
\begin{equation}
    \mathcal{Z}_{U(1)}=e^{-(m_1+m_2+\epsilon) \Lambda/2\epsilon_1\epsilon_2}\,.
\end{equation}
For reference, we give the one-instanton partition functions:
\begin{equation}
    \begin{aligned}
    & \mathcal{Z}_{U(2)}^{\mathrm{inst}}(\Lambda,a,m_1,m_2,m_3) = 1 + \frac{\prod_{i=1}^3\left(-a+m_i+\frac{\epsilon}{2}\right)}{2a\epsilon_1\epsilon_2(-2a+\epsilon)}\Lambda - \frac{\prod_{i=1}^3\left(a+m_i+\frac{\epsilon}{2}\right)}{2a\epsilon_1\epsilon_2(2a+\epsilon)}\Lambda + \mathcal{O}(\Lambda^2) \\
    & \mathcal{Z}_{SU(2)}^{\mathrm{inst}}(\Lambda,a,m_1,m_2,m_3) = 1 - \frac{\epsilon^2-4a^2-4m_1 m_2}{2\epsilon_1\epsilon_2(\epsilon+2a)(\epsilon-2a)}m_3\Lambda + \mathcal{O}(\Lambda^2)
    \end{aligned}
\end{equation}
\subsection{The Nekrasov-Shatashvili limit}
While the above formulae are valid for arbitrary $\epsilon_1,\epsilon_2$, in the context of the black hole we work in the Nekrasov-Shatashvili (NS) limit which is defined by $\epsilon_2\to0$ while keeping $\epsilon_1$ finite \cite{NEKRASOV_2010}. Furthermore we set $\epsilon_1=1$. The conformal blocks $\langle \Delta_\alpha, \Lambda_0, m_0 | V_{\alpha_{2}} (1) | \Delta_{\alpha_1} \rangle$ then need to be understood as being computed as a partition function in the NS limit. This is done by computing it for arbitrary $\epsilon_1,\epsilon_2$ and taking $\epsilon_1=1$ and $\epsilon_2\to0$ in the end, because the partition function itself diverges in this limit, while the ratios appearing e.g. in the connection formulas remain finite. Furthermore, we define the instanton part of the NS free energy as
\begin{equation}
    \mathcal{F}^{\mathrm{inst}}(\Lambda,a,m_1,m_2,m_3,\epsilon_1)=\epsilon_1 \lim_{\epsilon_2\to0}\epsilon_2 \log \mathcal{Z}^{\mathrm{inst}}_{SU(2)}(\Lambda,a,m_1,m_2,m_3,\epsilon_1,\epsilon_2)\,.
\end{equation}
One also uses the Matone relation \cite{Matone_1995}
\begin{equation}
    E=a^2-\Lambda \partial_\Lambda \mathcal{F}^{\mathrm{inst}}\,,
\end{equation}
which can be inverted order by order in $\Lambda$ to obtain $a(E)$. For reference, we give some relevant quantities computed up to one instanton, with $\epsilon_1=1$ and the leading power of $\epsilon_2$.
\begin{equation}
    \begin{aligned}
    & \mathcal{Z}^{\mathrm{inst}}_{SU(2)}(\Lambda,a,m_1,m_2,m_3) = 1-\frac{\frac{1}{4}-a^2-m_1 m_2}{\frac{1}{2}-2a^2}\frac{m_3 \Lambda}{\epsilon_2}+\mathcal{O}(\Lambda^2) \\
    & \mathcal{F}^{\mathrm{inst}}(\Lambda,a,m_1,m_2,m_3) = -\frac{\frac{1}{4}-a^2-m_1 m_2}{\frac{1}{2}-2a^2} m_3 \Lambda+\mathcal{O}(\Lambda^2) \\
    & a(E) = \sqrt{E} - \frac{\frac{1}{4}-E+a_1^2-a_2^2}{\sqrt{E}\left(1-4E\right)}m_3\Lambda+\mathcal{O}(\Lambda^2) \\
    & \frac{\mathcal{Z}^{\mathrm{inst}}_{SU(2)}(\Lambda,a,m_1-\frac{\theta'\epsilon_2}{2},m_2-\frac{\theta'\epsilon_2}{2},m_3)}{\mathcal{Z}^{\mathrm{inst}}_{SU(2)}(\Lambda,a,m_1-\frac{\theta\epsilon_2}{2},m_2+\frac{\theta\epsilon_2}{2},m_3)} = 1-\frac{\theta(m_1-m_2)+\theta'(m_1+m_2)}{1-4a^2}m_3 \Lambda+\mathcal{O}(\Lambda^2) \,.
    \end{aligned}
\end{equation}
Finally, we define the full NS free energy, including the classical and the one-loop part by
\begin{equation}
\begin{aligned}
\partial_a\mathcal{F}(\Lambda,a,m_1,m_2,m_3,\epsilon_1)=& -  2 a \log \frac{\Lambda}{\epsilon_1}  + 2 \epsilon_1 \log \frac{\Gamma \left(1 + \frac{2a}{\epsilon_1} \right) }{\Gamma \left(1 - \frac{2a}{\epsilon_1} \right)} + \epsilon_1 \sum_{i=1}^3 \log  \frac{\Gamma \left( \frac{1}{2} + \frac{m_i - a}{\epsilon_1} \right)}{\Gamma \left( \frac{1}{2} + \frac{m_i + a}{\epsilon_1} \right)} +\\ &+\partial_a\mathcal{F}^{\mathrm{inst}}(\Lambda,a,m_1,m_2,m_3,\epsilon_1) \,.
\end{aligned}
\end{equation}

\section{The semiclassical absorption coefficient}\label{appendix_semiclassical}
We give the detailed reduction of the full absorption coefficient in the semiclassical regime to the final result $\sigma = \exp -a_D/\epsilon_1$, with
\begin{equation}
    a_D := \oint_B \phi_{SW}(z)dz = \lim_{\epsilon_1\to0} \partial_a\mathcal{F}
\end{equation}
where $\phi_{SW}(z)$ is the Seiberg-Witten differential of the $\mathcal{N}=2$ $SU(2)$ gauge theory with three flavours and $\mathcal{F}$ is the full NS free energy introduced in the previous section.
 First we restore the powers of $\epsilon_1$ which were previously set to one in the exact absorption coefficient and substitute the AGT dictionary (see Appendix \ref{AppendixNekrasov}):
\begin{equation}
    \sigma = \frac{\displaystyle{-\mathrm{Im}\frac{m_1+m_2}{\epsilon_1}}}{\displaystyle{\left|\frac{\Gamma\left(1 + \frac{2a}{\epsilon_1}\right)\Gamma\left(\frac{2a}{\epsilon_1}\right) \Gamma\left(1 + \frac{m_1+m_2}{\epsilon_1}\right)\left(\frac{\Lambda}{\epsilon_1}\right)^{\frac{-a + m_3}{\epsilon_1}}}{\prod_{i=1}^3 \Gamma\left(\frac{1}{2}+\frac{m_i +a}{\epsilon_1}\right)} \frac{\mathcal{Z}^{\mathrm{inst}}_{SU(2)}\left(\Lambda,a+\frac{\epsilon_2}{2},m_1 ,m_2,m_3+\frac{\epsilon_2}{2}\right)}{\mathcal{Z}^{\mathrm{inst}}_{SU(2)}\left(\Lambda,a,m_1 -\frac{\epsilon_2}{2},m_2 -\frac{\epsilon_2}{2},m_3\right)} + (a \rightarrow -a) \right|^2}}
\end{equation}
In the regime where we have two real turning points and where we have obtained the semiclassical transmission coefficient we have $|\Lambda|\ll 1$. Then $a$ can be obtained order by order in an expansion in $\Lambda$, starting from $a = \ell + \frac{1}{2}+\mathcal{O}(\Lambda)$ by using the relation $E=a^2-\Lambda\partial_\Lambda \mathcal{F}^{\mathrm{inst}}$. Since $\Lambda\partial_\Lambda \mathcal{F}^{\mathrm{inst}}$ is real for real $a$, we see that all terms in the expansion and therefore $a$ itself are real. We anticipate that in $\sigma$ the term surviving in the semiclassical limit will be the first term in the denominator. This can be seen quickly by approximating $\Gamma(z)\approx e^{z\log z}$ for large $z$. In the semiclassical limit $a \gg m_i$ and the contribution of the five Gamma functions containing $a$ in the first term goes like $e^{\frac{a}{\epsilon_1}\log\frac{a}{\epsilon_1}}$. Extracting the term $\epsilon_1^{-\frac{a}{\epsilon_1}}$ cancels the explicit power of $\epsilon_1$ outside, and the rest of the exponential blows up. On the other hand the behaviour of the second term in the denominator of the transmission coefficient can be obtained by sending $a\rightarrow -a$, so we see that in this case the exponential vanishes, and indeed the dominant term  is the first one. The Gamma functions give the correct semiclassical one-loop contributions using Stirling's formula, and the ratio of partition functions gives the correct instanton contribution to $a_D$. 

More in detail, we can split the contributions to $a_D$ as
\begin{equation}
    a_D=a_D^{\mathrm{1-loop}} + a_D^{\mathrm{inst}} = a_{D,\mathrm{vector}}^{\mathrm{1-loop}} + \sum_{i=1}^3 a_{D,\mathrm{matter}}^{\mathrm{1-loop}} + a_{D,\mathrm{vector}}^{\mathrm{inst}}+ \sum_{i=1}^3 a_{D,\mathrm{matter}}^{\mathrm{inst}}\,.
\end{equation}
We take all matter multiplets to be in the antifundamental representation of $SU(2)$. The vector and matter multiplet one-loop contributions to $a_D$ are
\begin{equation}
    \begin{aligned}
    & a_{D,\mathrm{vector}}^{\mathrm{1-loop}}(a) = -8a+4a\log\frac{2a}{\Lambda}+4a\log\frac{-2a}{\Lambda}\\
    & a_{D,\mathrm{matter}}^{\mathrm{1-loop}}(a,m) = \left(a-m\right)\left[1-\log\left(\frac{-a+m}{\Lambda}\right)\right] +\left(a+m\right)\left[1-\log\left(\frac{a+m}{\Lambda}\right)\right] \,.
    \end{aligned}
\end{equation}
These are antisymmetric under $a\rightarrow -a$ as they should be. On the other hand, in the absorption coefficient we have several Gamma functions, which we can expand in the semiclassical limit using Stirling's approximation $\log\Gamma(z) = (z-1/2)\log z -z+ \mathcal{O}(z^{-1})$. We neglect the constant factors of $2\pi$ since we have the same amount of Gamma functions in the numerator and denominator and they will cancel. We have for the vectormultiplet
\begin{equation}
    \begin{aligned}
    \epsilon_1 \log\left|\Gamma\left(\frac{2a}{\epsilon_1}\right)\Gamma\left(1+\frac{2a}{\epsilon_1}\right)\right|^{-2} & \to 8a- 8a\log2a+8a\log\epsilon_1\\
    & = -a_{D,\mathrm{vector}}^{\mathrm{1-loop}}(a) - 4 \pi i a - 8a\log\frac{\Lambda}{\epsilon_1}\,,
    \end{aligned}
\end{equation}
for the matter multiplets
\begin{equation}
    \begin{aligned}
    \epsilon_1 \log\left|\Gamma\left(\frac{1}{2}+\frac{m+a}{\epsilon_1}\right)\right|^2 & \to (a-m)\log(a-m) + (a+m)\log(a+m)-2a(1+\log\epsilon_1)\\
    & = -a_{D,\mathrm{matter}}^{\mathrm{1-loop}}(a,m) + i\pi(a-m) + 2a \log\frac{\Lambda}{\epsilon_1}
    \end{aligned} 
\end{equation}
and there is one more Gamma function:
\begin{equation}
    \left|\Gamma\left(1+\frac{m_1+m_2}{\epsilon_1}\right)\right|^2 \to  i\frac{m_1+m_2}{\epsilon_1}e^{-i\pi (m_1+m_2)/\epsilon_1}.
\end{equation}
The last contribution is
\begin{equation}
    \left|\frac{\Lambda}{\epsilon_1}\right|^{2\frac{a - m_3}{\epsilon_1}} = \left(\frac{\Lambda}{\epsilon_1}\right)^{\frac{2a}{\epsilon_1}}e^{i\pi\frac{a + m_3}{\epsilon_1}}.
\end{equation}
Putting it all together we have
\begin{equation}
    -\mathrm{Im}\frac{m_1+m_2}{\epsilon_1}\left|\frac{\prod_{i=1}^3 \Gamma\left(\frac{1}{2}+\frac{m_i +a}{\epsilon_1}\right)\left(\frac{\Lambda}{\epsilon_1}\right)^{\frac{a - m_3}{\epsilon_1}}}{\Gamma\left(1+ \frac{2a}{\epsilon_1}\right)\Gamma\left(\frac{ 2a}{\epsilon_1}\right) \Gamma\left(1 + \frac{m_1+m_2}{\epsilon_1}\right)}\right|^2 \to e^{-a_D^{\mathrm{1-loop}}/\epsilon_1}\,.
\end{equation}
Now let us look at the instanton partition functions:
\begin{equation}
    \begin{aligned}
    \frac{\mathcal{Z}^{\mathrm{inst}}_{SU(2)}\left(\Lambda,a,m_1 -\frac{\epsilon_2}{2},m_2 -\frac{\epsilon_2}{2},m_3\right)}{\mathcal{Z}^{\mathrm{inst}}_{SU(2)}\left(\Lambda,a+\frac{\epsilon_2}{2},m_1 ,m_2,m_3+\frac{\epsilon_2}{2}\right)} =  \frac{\mathcal{Z}^{\mathrm{inst}}_{SU(2)}\left(\Lambda,\tilde{a}-\frac{\epsilon_2}{2},m_1 -\frac{\epsilon_2}{2},m_2 -\frac{\epsilon_2}{2},\tilde{m}_3-\frac{\epsilon_2}{2}\right)}{\mathcal{Z}^{\mathrm{inst}}_{SU(2)}\left(\Lambda,\tilde{a},m_1 ,m_2,\tilde{m}_3\right)}
    \end{aligned}
\end{equation}
where we have defined $\tilde{a} = a+\frac{\epsilon_2}{2}$ and $\tilde{m}_3 = m_3+ \frac{\epsilon_2}{2}$. Now, looking at the explicit expressions for the $U(2)$ Nekrasov partition functions, we see that the part corresponding to the gauge field depends only on $a_1 - a_2$, and the part corresponding the the hypermultiplets only on $a_1 + m_i$ and $a_2 + m_i$. So we see that
\begin{equation}
    \begin{aligned}
    & \mathcal{Z}^{\mathrm{inst}}_{U(2)}\left(\Lambda,a_1=\tilde{a}-\frac{\epsilon_2}{2},a_2 = -\tilde{a}+\frac{\epsilon_2}{2},m_1 -\frac{\epsilon_2}{2},m_2 -\frac{\epsilon_2}{2},\tilde{m}_3-\frac{\epsilon_2}{2}\right)  = \\ = \,& \mathcal{Z}^{\mathrm{inst}}_{U(2)}\left(\Lambda,a_1=\tilde{a}-\epsilon_2,a_2 = -\tilde{a},m_1 ,m_2 ,\tilde{m}_3\right)\,.
    \end{aligned}
\end{equation}
The $U(1)$ part behaves as
\begin{equation}
    \begin{aligned}
    \mathcal{Z}^{-1}_{U(1)}(\Lambda,m_1-\frac{\epsilon_2}{2},m_2-\frac{\epsilon_2}{2}) = e^{(m_1+m_2-\epsilon_2)\Lambda/2\epsilon_1\epsilon_2} =e^{-\Lambda/2\epsilon_1} \mathcal{Z}^{-1}_{U(1)}(\Lambda,m_1,m_2)\,,
    \end{aligned}
\end{equation}
therefore
\begin{equation}
    \begin{aligned}
    &\frac{\mathcal{Z}^{\mathrm{inst}}_{SU(2)}\left(\Lambda,a,m_1 -\frac{\epsilon_2}{2},m_2 -\frac{\epsilon_2}{2},m_3\right)}{\mathcal{Z}^{\mathrm{inst}}_{SU(2)}\left(\Lambda,a+\frac{\epsilon_2}{2},m_1 ,m_2,m_3+\frac{\epsilon_2}{2}\right)} =e^{-\Lambda/2\epsilon_1} \frac{\mathcal{Z}^{\mathrm{inst}}_{SU(2)}\left(\Lambda,\tilde{a}-\epsilon_2,-\tilde{a},m_1 ,m_2 ,\tilde{m}_3\right)}{\mathcal{Z}^{\mathrm{inst}}_{SU(2)}\left(\Lambda,\tilde{a},-\tilde{a},m_1,m_2,\tilde{m}_3\right)} =\\
    & =e^{-\Lambda/2\epsilon_1} \exp\frac{1}{\epsilon_1 \epsilon_2}\left\{\mathcal{F}^{\mathrm{inst}}\left(\Lambda,\tilde{a}-\epsilon_2,-\tilde{a},m_1 ,m_2 ,\tilde{m}_3\right) - \mathcal{F}^{\mathrm{inst}}\left(\Lambda,\tilde{a},-\tilde{a},m_1,m_2,\tilde{m}_3\right)\right\} = \\
    & = e^{-\Lambda/2\epsilon_1}\exp -\frac{1}{\epsilon_1} \frac{\partial}{\partial a_1} \mathcal{F}^{\mathrm{inst}}\left(\Lambda,a_1,a_2,m_1,m_2,\tilde{m}_3\right)|_{a_1 = \tilde{a},\,a_2=-\tilde{a}}\,.
    \end{aligned}
\end{equation}
Now there are no more factors of $1/\epsilon_2$ so we can safely drop the tildes. On the other hand, by symmetry considerations which are most easily seen in the expression as a conformal block, and using the fact that $\Lambda$ and the three masses are purely imaginary while $a$ is real, we have
\begin{equation}
    \begin{aligned}
    &\frac{\mathcal{Z}^{\mathrm{inst}}_{SU(2)}\left(\Lambda,a,m_1 -\frac{\epsilon_2}{2},m_2 -\frac{\epsilon_2}{2},m_3\right)}{\mathcal{Z}^{\mathrm{inst}}_{SU(2)}\left(\Lambda,a+\frac{\epsilon_2}{2},m_1 ,m_2,m_3+\frac{\epsilon_2}{2}\right)} = \frac{\mathcal{Z}^{\mathrm{inst}}_{SU(2)}\left(-\Lambda,a,-m_1 +\frac{\epsilon_2}{2},-m_2 +\frac{\epsilon_2}{2},-m_3\right)}{\mathcal{Z}^{\mathrm{inst}}_{SU(2)}\left(-\Lambda,a+\frac{\epsilon_2}{2},-m_1 ,-m_2,-m_3-\frac{\epsilon_2}{2}\right)} =\\ & =\frac{\mathcal{Z}^{\mathrm{inst}}_{SU(2)}\left(\Lambda^*,a^*,m_1^* +\frac{\epsilon_2}{2},m_2^* +\frac{\epsilon_2}{2},m_3^*\right)}{\mathcal{Z}^{\mathrm{inst}}_{SU(2)}\left(\Lambda^*,a^*+\frac{\epsilon_2}{2},m_1^*,m_2^*,m_3^*-\frac{\epsilon_2}{2}\right)}
    \end{aligned}
\end{equation}
And therefore, repeating the same steps as above,
\begin{equation}
    \left(\frac{\mathcal{Z}^{\mathrm{inst}}_{SU(2)}\left(\Lambda,a,m_1 -\frac{\epsilon_2}{2},m_2 -\frac{\epsilon_2}{2},m_3\right)}{\mathcal{Z}^{\mathrm{inst}}_{SU(2)}\left(\Lambda,a+\frac{\epsilon_2}{2},m_1 ,m_2,m_3+\frac{\epsilon_2}{2}\right)}\right)^* = e^{\Lambda/2\epsilon_1}\exp \frac{1}{\epsilon_1} \frac{\partial}{\partial a_2} \mathcal{F}^{\mathrm{inst}}\left(\Lambda,a_1,a_2,m_1,m_2,m_3\right)|_{a_1 =a,\,a_2=-a}
\end{equation}
Now using $\partial_a \mathcal{F}=\partial_{a_1} \mathcal{F}-\partial_{a_2} \mathcal{F}$ we have: 
\begin{equation}
    \left|\frac{\mathcal{Z}^{\mathrm{inst}}_{SU(2)}\left(\Lambda,a,m_1 -\frac{\epsilon_2}{2},m_2 -\frac{\epsilon_2}{2},m_3\right)}{\mathcal{Z}^{\mathrm{inst}}_{SU(2)}\left(\Lambda,a+\frac{\epsilon_2}{2},m_1 ,m_2,m_3+\frac{\epsilon_2}{2}\right)}\right|^2 = e^{-a_D^{\mathrm{inst}}/\epsilon_1}
\end{equation}
which combined with the one-loop part finally gives
\begin{equation}
    \resizebox{0.14\hsize}{!}{\boxed{\sigma \approx e^{-a_D/\epsilon_1}}}\,.
\end{equation}
This result is valid for $\ell \gg 1$ and $M\omega,\text{a}\omega \ll 1$, while keeping all orders in $M\omega,\text{a}\omega$.

\printbibliography[heading=bibintoc]
\end{document}